\documentclass[11pt]{article}
\usepackage{latexsym,cite,amssymb,amsmath,times}
\usepackage{enumitem}
\usepackage{float}
\usepackage{xcolor}
 
\usepackage{framed}

\usepackage{accents}
\newcommand*\udot[1]{%
  \underaccent{\dot}{#1}}

\makeatletter

\@addtoreset{equation}{section} \makeatother

\input amssym.def
\input amssym.tex

\newcommand{\half}{{{\textstyle\frac{1}{2}}}}
\newcommand{\quarter}{{{\textstyle\frac{1}{4}}}}
\newcommand{\be}{\begin{equation}}
\newcommand{\ee}{\end{equation} }
\newcommand{\beqa}{\begin{eqnarray} }
\newcommand{\eeqa}{\end{eqnarray} }
\newcommand{\ba}{\begin{array}}
\newcommand{\ea}{\end{array}}

\newcommand{\so}{\mathbf{so}}

\newcommand{\Spin}{\mathbf{Spin}}

\newcommand{\HS}{\mathbf{HS}}

\newcommand{\rmC}{{\rm C}}

\newcommand{\ODD}{\mathbf{O}(D,D)}

\newcommand{\Off}{\mathbf{O}(4,4)}
\newcommand{\Spinf}{{\Spin(1,3)}}
\newcommand{\oSpinf}{{{\Spin}(3,1)}}
\newcommand{\HSf}{{\HS(4)}}

\newcommand{\YM}{\scriptscriptstyle{\rm{YM}}}
\newcommand{\Vasiliev}{\scriptscriptstyle{\rm{def.osc.}}}

\newcommand\Tr{{\rm Tr}}

\newcommand\rd{{\rm d}}

\newcommand\bfk{\mathbf{k}}
\newcommand\brbfk{\mathbf{\bar{k}}}

\newcommand\cD{{\cal D}}

\newcommand\cF{{\cal F}}

\newcommand\cJ{{\cal J}}

\newcommand\cL{{\cal L}}

\newcommand\cP{{\cal P}}
\newcommand\cQ{{\cal Q}}
\newcommand\cR{{\cal R}}
\newcommand\cS{{\cal S}}
\newcommand\cT{{\cal T}}

\newcommand\cW{{\cal W}}

\newcommand\brcP{{\bar{\cP}}}

\newcommand\hcL{{\hat{\cal L}}}
\newcommand\hcO{{\hat{\cal O}}}

\newcommand\dalpha{{\dot{\alpha}}}
\newcommand\dbeta{{\dot{\beta}}}
\newcommand\dgamma{{\dot{\gamma}}}

\newcommand\udalpha{{\udot{\alpha}}}
\newcommand\udbeta{{\udot{\beta}}}

\newcommand\uddelta{\udot{\delta}}

\newcommand\hzeta{\hat{\zeta}}
\newcommand\hbrzeta{\hat{\brzeta}}

\newcommand\hP{\hat{P}{}}

\newcommand\hM{\hat{M}{}}

\newcommand\brzeta{\bar{\zeta}}

\newcommand\gHS{g_{\rm \scriptscriptstyle{HS}}}
\newcommand\sHS{{\scriptscriptstyle{\rm HS}}}
\newcommand\HSDFT{{\scriptscriptstyle{\rm HS-DFT}}}
\newcommand\DFT{{\scriptscriptstyle{\rm DFT}}}

\newcommand\dis{\displaystyle}

\def\tx{\tilde{x}}

\def\brepsilon{\bar{\epsilon}}
\def\brrho{\bar{\rho}}

\def\brp{{\bar{p}}}

\def\brs{{\bar{s}}}
\def\bry{{\bar{y}}}
\def\brz{{\bar{z}}}

\def\brV{{\bar{V}}}
\def\brP{{\bar{P}}}

\newcommand{\na}{{\nabla}}
\newcommand{\trd}{{\bigtriangledown}}

\newcommand{\gammaf}{\gamma^{(5)}}

\def\lpartial{{\stackrel{\leftarrow}{\partial}}}
\def\rpartial{{\stackrel{\rightarrow}{\partial}}}

\newcommand\mr{\mathring}

\begin{document}
\begin{titlepage}
\title{
\vskip 2cm
Higher Spin Double Field Theory\,: A Proposal
~\\
~\\}
\author{\sc   Xavier Bekaert${}^{\sharp}$\mbox{~}   and \mbox{~} 
Jeong-Hyuck Park${}^{\dagger}$ }
\date{}
\maketitle \vspace{-1.0cm}
\begin{center}
~~~\\
${}^{\sharp}$Laboratoire de Math\'ematiques et Physique Th\'eorique\\
    Unit\'e Mixte de Recherche 7350 du CNRS, F\'ed\'eration Denis Poisson\\
    Universit\'e Fran\c cois Rabelais, 
    Parc de Grandmont, 37200 Tours, France\\

\vspace{2mm}
${}^{\dagger}$Department of Physics, Sogang University, 35 Baekbeom-ro, Mapo-gu, Seoul 04107, Korea\\

\vspace{2mm}
 ${}^{\sharp,\dagger}$B.W. Lee Center for Fields, Gravity and Strings, Institute for Basic Science, Daejeon 34047, Korea\\
~\\
~{}\\
\texttt{bekaert@lmpt.univ-tours.fr\qquad park@sogang.ac.kr}
~~~\\~\\~\\
\end{center}
\begin{abstract}
\vskip0.2cm
\noindent
We construct a double field theory coupled to the fields present in Vasiliev's equations. Employing  the ``semi-covariant" differential geometry, we spell a functional in which each term is completely covariant with respect to $\mathbf{O}(4,4)$  T-duality,  doubled diffeomorphisms,   $\mathbf{Spin}(1,3)$    local Lorentz symmetry  and, separately,  $\mathbf{HS}(4)$  higher spin gauge symmetry. We  identify   a minimal set of BPS-like conditions whose solutions automatically satisfy  the full  Euler-Lagrange equations. As such a solution, we derive  a linear dilaton  vacuum.  With extra algebraic constraints further supplemented, the BPS-like conditions reduce to the bosonic  Vasiliev  equations.
\end{abstract}

\thispagestyle{empty}
\end{titlepage}
\newpage
\tableofcontents 



~\\
~\\
~\\

\section{Introduction}

Both higher spin gravity and double field theory extend Einstein's gravity beyond the Riemannian paradigm. The idea of unifying  the graviton  either with  an infinite tower of massless fields of ever higher spin, or with other massless NS-NS fields,  appears to  require us to extend  Riemannian geometry in both cases. \\

While free Higher Spin (HS)  theories were well explored, notably by Fronsdal \cite{Fronsdal:1978rb,Fronsdal:1978vb}, it was Vasiliev who first managed to write down a set of gauge invariant equations which describe the full interactions among the infinite tower of higher spin fields~\cite{Vasiliev:1992av} 
(see also \cite{Vasiliev:1995dn,Vasiliev:1999ba,Bekaert:2005vh,Didenko:2014dwa,Vasiliev:2014vwa,Vasiliev:2016sgg} for  reviews). The higher spin gauge symmetries are expected to be so huge as to prevent quantum corrections and thereby ensure ultraviolet finiteness of higher spin gravity at quantum level (see \textit{e.g.} \cite{Fradkin:1990kr} for an early statement of this conjecture).
Moreover, the higher spin symmetry enhancement of string theory in the ultra-Planckian regime has been argued to be responsible for the remarkable softness of its scattering amplitudes~\cite{Gross:1987kza}.
The holographic duality provides a complementary perspective on these expectations. On the one hand, interactions of massless higher spin particles suffer from serious restrictions in flat spacetime (see \textit{e.g.}~\cite{Bekaert:2010hw,Rahman:2013sta} for reviews on no-go theorems) and, in fact, Vasiliev equations require a nonvanishing cosmological constant and possess (anti) de Sitter spacetime as the most symmetric background. On the other hand, the holographic dictionary suggests that  higher spin gravity theories around anti de Sitter spacetime (\textit{AdS}) are dual to free or integrable conformal field theories on the boundary (see \textit{e.g.}~\cite{Bianchi:2005yh,Bekaert:2012ux,Giombi:2012ms} for some reviews on higher spin holography). Since what replaces the S-matrix in \textit{AdS} are boundary correlators, the remarkable simplicity of such holograms (free or integrable CFTs) is the avatar of the absence of quantum corrections and the extreme softness of higher spin gravity. While the  holographic duality suggests a relation between higher spin gravity and \textit{closed} string theory, there exists  a striking resemblance between the classification of allowed internal symmetry group extensions in Vasiliev theory and Chan-Paton factors in \textit{open} string theory.
Accordingly, the Vasiliev system has also been argued (see \textit{e.g.}~the review \cite{Sagnotti:2013bha} and references  therein) to yield an effective description for the first Regge trajectory of the \textit{open} string in the tensionless limit. Let us emphasize that, in the latter interpretation, the Vasiliev multiplet appears as some sort of 'matter' sector and it becomes questionable whether the massless spin-two particles in the Vasiliev multiplet should be treated as the graviton.
\\
 
Double Field Theory (DFT) is an extension of the Einstein gravity  in accordance with, especially,  string theory~\cite{Siegel:1993xq,Siegel:1993th,Hull:2009mi,Hull:2009zb,Hohm:2010jy,Hohm:2010pp}.  The primary  goal of DFT was to manifest T-duality, or to reformulate supergravity in a way that the hidden $\ODD$ structure   becomes manifest (see \textit{e.g.}~\cite{Aldazabal:2013sca,Berman:2013eva,Hohm:2013bwa} for reviews). As the $\ODD$ T-duality rotations mix the  Riemannian metric, the Kalb-Ramond $B$-field and the scalar dilaton,  the geometrization of the  whole massless NS-NS sector was inevitable in DFT, as  once anticipated by Siegel~\cite{Siegel:1993th}. This is clearly  in contrast to the conventional picture  of  (super)gravity,  where the Riemannian  metric is regarded as the only geometric object while  the $B$-field and the scalar dilaton are viewed as `matter' living  on the background the  metric characterizes. 
\\

Concretely in \cite{Jeon:2010rw,Jeon:2011cn},  the DFT generalization of  the Christoffel symbols was derived which is compatible with, and hence comprised of, the whole massless  NS-NS sector. It defines so-called the  `semi-covariant'  derivative and curvature which can be completely covariantized~\cite{Jeon:2010rw,Jeon:2011cn,Jeon:2011kp,Jeon:2011vx,Jeon:2012kd}.   
This line of geometrical  development, often called the `semi-covariant geometry', has turned out to have the potential   to replace the conventional Riemannian geometry, while  it manifests, for every single term in formulas,    $\ODD$ T-duality,   twofold local Lorentz symmetries, \textit{i.e.~}$\Spin(1,D{-1})\times\Spin(D{-1},1)$,  and  doubled diffeomorphisms (DFT-diffeomorphisms) which unifies the ordinary Riemannian  diffeomorphisms and the $B$-field gauge symmetry  (\textit{c.f.~}`Generalized Geometry'~\cite{Hitchin:2004ut,Hitchin:2010qz,Gualtieri:2003dx}). For example, for $D{=10}$, the half-maximal as well as the maximal supersymmetric double field theories have been constructed to the full order in fermions~\cite{Jeon:2011sq,Jeon:2012hp}, of which the latter unifies IIA and IIB supergravities since the twofold spin groups remove intrinsically the relative chirality difference of IIA and IIB. Further, as for $D{=4}$, it has been shown  possible to double field theorize the Standard Model, without any extra physical degree introduced~\cite{Choi:2015bga}:  equipped with the semi-covariant geometry, 
it can couple to an arbitrary NS-NS background in a completely covariant manner. In particular, the incorporation of the  twofold spin groups,   $\Spinf{\times\oSpinf}$, into the Standard Model lead to an  experimentally testable   prediction that the quarks and the leptons  may belong to the  different spin classes. The semi-covariant differential geometry also facilitates    efficient  perturbation analyses as well as  Wald type Noether charge derivations  in double field  theories~\cite{Ko:2015rha,Blair:2015eba,Park:2015bza}.\footnote{For other approaches, we refer to  \textit{e.g.~}\cite{Hohm:2010xe,Hohm:2011zr,Berman:2013uda,Hohm:2015ugy}.}\\

It is the purpose of the present paper to  apply the semi-covariant geometry  and propose a ``higher spin double field theory''. To be more precise, we extend DFT by introducing the fields present in Vasiliev equations which we treat as `matter'  minimally coupled to the  massless NS-NS sector, \textit{i.e.~}the geometric objects in DFT.\footnote{In our proposal, DFT is the genuine gravity theory (closed string related) while the HS sector is treated as matter (possibly open string related). The geometrization of the entire higher spin degrees of freedom  is of course desirable but it goes beyond the scope of the present work.}
Accordingly, in contrast to the pure Vasiliev equations, we shall have the notion of covariant derivatives which accompany the DFT-Christoffel connection as well as the local DFT-Lorentz spin connection.  For concreteness, putting ${D=4}$, we restrict ourselves to the spacetime dimension four. We focus on one of the twofold spin groups in DFT  and consider its extension  to include the higher spin gauge symmetry, $\HSf$,
\be
\ba{lll}
\Spinf~~~&\longrightarrow&~~~\Spinf\times\HSf\,.
\ea
\label{extension}
\ee
The algebra denoted here $\HSf$ is the algebra of gauge symmetries of Vasiliev equations.\footnote{This algebra was referred to as ``embedding algebra'' in \cite{Alkalaev:2014nsa}. It must be distinguished from what is usually referred to as ``higher spin algebra'' and sometimes denoted as $\mathfrak{hs}(4)$ in the literature. In fact, as is well known, the latter algebra is just a subalgebra of the former, $\mathfrak{hs}(4)\subset \HSf$, because it only depends on half of the oscillators. Nevertheless, in the present paper we will often loosely refer to $\HSf$ as the higher spin gauge algebra because here we never particularize to its subalgebra, $\mathfrak{hs}(4)$.}
In our proposal,  $\Spinf$ and $\HSf$ are realized independently,  although they share the same spinorial indices. This is analogous to the case of DFT where the  T-duality and doubled  diffeomorphisms act on the same indices, while they differ from each other.\\

We  present a functional  and derive the corresponding Euler-Lagrange equations. Every term in our formulas  is going to be   
completely covariant under     $\mathbf{O}(4,4)$ T-duality,  DFT-diffeomorphisms,     $\mathbf{Spin}(1,3)$ local Lorentz symmetry and    higher spin gauge symmetry, $\mathbf{HS}(4)$. We identify  a minimal set of BPS-like  conditions which  are strong enough to solve automatically the full equations of motion.  We also derive a   vacuum configuration as a solution to the BPS-like conditions, which  is characterized by a linear DFT-dilaton and a constant or flat DFT-vielbein.  Finally, we discuss a consistent truncation of  the BPS-like conditions to  the bosonic  Vasiliev  equations after  imposing extra algebraic  conditions. We also comment on   a possible alternative without any extra condition being imposed,    in view of the open string theory interpretation of Ref.\cite{Sagnotti:2013bha}.\\

Let us clarify why, as the attentive reader may have noticed, in the abstract and introduction, we have refrained from calling the invariant functional which we propose, an ``action''. This terminological caution aims at avoiding possible misunderstandings since no standard action principle is presently known for the precise Vasiliev equations.\footnote{However, a nonstandard proposal is available (\textit{c.f.}~the review \cite{Arias:2016ajh} and references therein for details). See also \cite{Vasiliev:2015mka} for a proposal of an on-shell action and related discussion.} The functional we propose is merely an extension of DFT where two fields taking values in the $\mathbf{HS}(4)$ algebra, which can be off-shell identified as the fields present in Vasiliev equations,  are added and treated as some sort of `matter' minimally coupled to the pure DFT (and hence possibly the open string interpretation). In fact, we want to stress that the equations of motion we obtain from the   functional are \textit{not} by themselves equal to the Vasiliev equations. Nevertheless, the latter can be obtained as a consistent truncation of the former. To be more precise, the differential equations (the flatness and covariant constancy) in Vasiliev system of equations are identified  in the present paper as part of the BPS-like conditions, while the algebraic equations, including the so-called deformed oscillator algebra, in Vasiliev system appear as a subclass of the  solutions  to the remaining BPS-like conditions.  They are not the most general solutions and hence, strictly speaking, must  be supplemented by hand if we want to precisely recover  the full  Vasiliev equations.\footnote{Off-shell, the two $\HSf$-valued fields are simply a gauge potential and a bosonic spinor field taking values in the infinite-dimensional algebra. Once the BPS-like conditions are imposed, then the gauge potential is flat (up to projection) and the spinor field is covariantly constant. Strictly speaking, it is only when the extra algebraic equations are eventually imposed that they can be interpreted as  the {Vasiliev's}  HS gauge fields.}
This word of caution being said, the BPS truncation of the Euler-Lagrange equations of our proposed functional  provides  a suggestive  double field theory covariantization of the Vasiliev bosonic equations.\footnote{Since the Vasiliev fields are here treated as ``matter'' minimally coupled to DFT, the massless spin-two field in the HS tower, if any, should be distinguished from the metric in the DFT multiplet. When the Vasiliev theory is thought as arising from the first Regge trajectory of open string in some tensionless limit (a point of view followed here), then
the massless spin-two particle in the higher spin  multiplet should indeed \textit{not} be identified with the graviton.
For this reason, we avoided the use of the term ``higher spin gravity'' when referring to our proposal.
}\\

The  rest of the paper is organized as follows. 
\begin{itemize}
\item In section~\ref{SECPROPOSAL},  we summarize our proposal of Higher Spin Double Field Theory, which decomposes into two parts:  kinematics and dynamics.  
\item Section~\ref{SECEXPOSITION} delivers detailed exposition. In a self-contained manner,  we review  $\Spinf$ Clifford algebra (gamma matrices,  Majorana condition),  the Wick (\textit{i.e.} normal) ordered star product, and the semi-covariant geometry of double field theory including its complete covariantization.  We detail the derivation of the  Euler-Lagrange equations as well as the BPS-like conditions.  We discuss  the linear dilaton vacuum solution, and explain  the consistent truncation of the BPS-like conditions  to the bosonic Vasiliev equations. 
\item We conclude with comments in section~\ref{SECCOMMENTS}. 
\end{itemize}

\section{Proposal\label{SECPROPOSAL}}
In this section, we depict the  salient features of the  Higher Spin Double Field Theory (HS-DFT) we propose.   We first state   the kinematic  ingredients, such as the symmetries, coordinates and the field content.  We then  introduce    the master  derivative, in terms of which  we spell the  proposed action, its  full Euler-Lagrange equations  and the BPS-like conditions. We  sketch   a linear dilaton vacuum solution,  and  prescribes  the consistent truncation of the BPS-like equations to the undoubled   Vasiliev equations.  The detailed exposition will  follow  in the next section~\ref{SECEXPOSITION}.\\

\subsection{Kinematics of HS-DFT\label{SECing}}

\begin{itemize}
\item First of all, the symmetries of the proposed HS-DFT  are the following. 
\begin{itemize}
\item  $\Off$ \textit{T-duality}, 
\item \textit{DFT-diffemorphisms},
\item $\Spinf$ \textit{local Lorentz symmetry},
\item $\HSf$ \textit{higher spin gauge symmetry}.\\

\end{itemize}

\item The relevant indices are  summarized in Table~\ref{TABindices}  as for  our convention.

{{\begin{table}[H]
\begin{center}
\begin{tabular}{c|c|c}
Index~&~Representation~&~Range~\\
\hline
$A,B,\cdots$ & ~~\begin{tabular}{c}$\Off$ vector\\ ~~DFT-diffeomorphisms\end{tabular}
~&~$1,2,\cdots 8$~~\quad\\
\hline
$\alpha,\beta,\cdots$&~~\begin{tabular}{c}$\Spinf$ spinor~ \\
$\HSf$ spinor  \end{tabular} &~$1,2,3,4$~\\
\hline
$p,q,\cdots$&\quad~~$\Spinf$  vector\quad~&~$0,1,2,3$~ \\
\end{tabular}
\caption{Index  for  each symmetry representation. As is always the case with  DFT, 
$\Off$  and DFT-diffeomorphisms share the same vectorial indices, \textit{i.e.~}the capital Latin alphabet letters. Similarly,  $\Spinf$ and   $\HSf$ use the same spinorial indices, \textit{i.e.~}the small greek letters. In our proposal of HS-DFT, $\,\Spinf$ is not a subgroup of $\HSf$, \textit{c.f.~}(\ref{HSGS}) and (\ref{Spinftransf}).} 
\label{TABindices}
\end{center}
\end{table}}}
In particular, the $\Off$  group is characterized by the  constant  invariant  metric, $\cJ_{AB}$,  put   in an off block diagonal form,
\be
\ba{lll}
L_{A}{}^{B}\in\Off~~:&\qquad
L_{A}{}^{C}L_{B}{}^{D}\cJ_{CD}=\cJ_{AB}\,,\quad&\quad
\cJ_{AB}=\left(\ba{cc}0&1\\1&0\ea\right)\,.
\ea
\label{cJ}
\ee
The capital Latin alphabet  indices are used commonly for both $\Off$ and DFT-diffeomorphisms, and they can be freely lowered or raised by the above  invariant metric, or its inverse, $\cJ^{AB}$.

The $\Spin(1,3)$ vectorial, small Latin alphabet indices are subject to the mostly positive  Minkowskian metric,  
$\eta_{pq}$, which also sets  the Clifford algebra of the  $4\times 4$ gamma matrices,
$(\gamma^{p})^{\alpha}{}_{\beta}$, 
\be
\ba{lll}
\gamma^{p}\gamma^{q}+\gamma^{q}\gamma^{p}=2\eta^{pq}\,,\quad&\quad
\eta_{pq}=\mbox{diag}(-+++)\,,\quad&\quad \gammaf=(\gammaf)^{\dagger}=(\gammaf)^{-1}=-i\gamma^{0123}\,.
\ea
\label{gammabr}
\ee  
We set a  unitary and Hermitian matrix, $\mathbf{A}=\mathbf{A}^{\dagger}=\mathbf{A}^{-1}\equiv-i\gamma^{0}$, satisfying for $n=0,1,2,3,4$,
\be 
\ba{lll}
(\gamma^{p})^{\dagger}=\gamma_{p}=-\mathbf{A}\gamma^{p}\mathbf{A}^{-1}\,,~&
(\mathbf{A}\gamma^{p_{1}p_{2}\cdots p_{n}})^{\dagger}=(-1)^{n(n+1)/2}\mathbf{A}
\gamma^{p_{1}p_{2}\cdots p_{n}}\,,~&
\gammaf=-\mathbf{A}\gammaf\mathbf{A}^{-1}\,.
\ea
\label{mathbfA}
\ee
The  charge conjugation matrix is  skew-symmetric and unitary,  $\mathbf{C}=-\mathbf{C}^{T}=(\mathbf{C}^{\dagger})^{-1}$, satisfying 
\be
\ba{lll}
(\gamma^{p})^{T}=-\mathbf{C}\gamma^{p}\mathbf{C}^{-1}\,,&
(\mathbf{C}\gamma^{p_{1}p_{2}\cdots p_{n}})_{\alpha\beta}=-(-1)^{n(n+1)/2}(\mathbf{C}\gamma^{p_{1}p_{2}\cdots p_{n}})_{\beta\alpha}\,,&(\gammaf)^{T}=\mathbf{C}\gammaf \mathbf{C}^{-1}\,.
\ea
\label{mathbfC}
\ee
Especially, $\mathbf{C}\gamma^{p}$ and $\mathbf{C}\gamma^{pq}$  are symmetric to fulfill   the  completeness relation,
\be
\half(\delta_{\alpha}^{\,~\gamma}\delta_{\beta}^{\,~\delta}
+\delta_{\beta}^{\,~\gamma}\delta_{\alpha}^{\,~\delta})=
\quarter(\mathbf{C}\gamma_{p})_{\alpha\beta}(\gamma^{p}\mathbf{C}^{-1})^{\gamma\delta}-\textstyle{\frac{1}{8}}(\mathbf{C}\gamma_{pq})_{\alpha\beta}(\gamma^{pq}\mathbf{C}^{-1})^{\gamma\delta}\,.
\label{complete}
\ee
On the other hand,  $\mathbf{C}$, $\mathbf{C}\gammaf$ and $\mathbf{C}\gammaf\gamma^{p}$  are  skew-symmetric,  with the completeness relation,
\be
\half(\delta_{\alpha}^{\,~\gamma}\delta_{\beta}^{\,~\delta}
-\delta_{\beta}^{\,~\gamma}\delta_{\alpha}^{\,~\delta})=
-\quarter \mathbf{C}_{\alpha\beta}\mathbf{C}^{-1\gamma\delta}
-\quarter(\mathbf{C}\gammaf)_{\alpha\beta}(\gammaf \mathbf{C}^{-1})^{\gamma\delta}
-\quarter(\mathbf{C}\gammaf\gamma^{p})_{\alpha\beta}(\gamma_{p}\gammaf \mathbf{C}^{-1})^{\gamma\delta}\,.
\label{complete2}
\ee

\item 
The four-dimensional  spacetime is described  by adopting the  eight-dimensional   \textit{doubled-yet-gauged}  coordinate system, $\{x^{A}\}$, $A=1,2,\cdots,8$~\cite{Park:2013mpa,Lee:2013hma}.  This means that -- as reviewed in section~\ref{SECDyG}  -- all the  variables, \textit{e.g.~}fields, local parameters, their arbitrary derivatives and  products,  are subject to  the \textit{section condition}~\cite{Hohm:2010pp},
\be
\partial_{A}\partial^{A}= 0\,.
\label{seccon}
\ee
Consequently,  the theory is confined to live on a four-dimensional `section'. 
The diffeomorphisms in such a doubled-yet-gauged coordinate system are generated by the generalized Lie derivative used in DFT, \textit{c.f.}~(\ref{GLieDeriv}).

\item The higher spin  gauge  symmetry, $\HSf$, is realized through  a star product defined over an internal  space with $\Spinf$ spinorial coordinates, $\zeta^{\alpha}$ and $\brzeta_{\beta}$,
\be 
\ba{lll}
[\zeta^{\alpha},\brzeta_{\beta}]_{\star}=\delta^{\alpha}_{~\beta}\,,\quad&\quad
[\zeta^{\alpha},\zeta^{\beta}]_{\star}=0\,,\quad&\quad
[\brzeta_{\alpha},\brzeta_{\beta}]_{\star}=0\,.
\ea
\label{zetacom}
\ee
In particular, the spinorial coordinates are  bosonic (\textit{i.e.~}even Grassmannian)     Dirac spinors having four  complex components, and $\brzeta_{\alpha}$ is the  Dirac conjugate of $\zeta^{\alpha}$, \textit{i.e.}   $\brzeta\equiv\zeta^{\dagger}\mathbf{A}\,$.

As is the case with Vasiliev~\cite{Vasiliev:1992av}, the star product represents  a non-commutative   algebra which is  not Weyl but Wick ordered, and hence generically it reads
\be
f(x,\zeta,\brzeta)\star g(x,\zeta,\brzeta)=\dis{f(x,\zeta,\brzeta)~\exp\left(
\frac{\lpartial~~}{\partial{\zeta}^{\alpha}}
\frac{\rpartial~~}{\partial \brzeta_{\alpha}}\right)g(x,\zeta,\brzeta)}\,.
\label{starproduct}
\ee
This presentation of the algebra of gauge symmetries in Vasiliev equations is not standard. The detailed relation to the more traditional $Y$ and $Z$ oscillators is given  in section~\ref{trunc}.
~\\

\item \textit{The  field content} of HS-DFT is
\be
\ba{llll}
d(x)\,,\quad&\quad V_{Ap}(x)\,,\quad&\quad \cW_{A}(x,\zeta,\brzeta)\,,\quad&\quad
\Psi^{\alpha}(x,\zeta,\brzeta)\,,
\ea
\label{FC}
\ee
which correspond  to  a DFT-dilaton, a DFT-vielbein, a higher spin gauge potential and a bosonic $\Spinf$ Majorana spinor.  In particular, the DFT-dilaton gives rise to a scalar density with weight one after exponentiation, $e^{-2d}$, and the DFT-vielbein is subject to one defining property, 
\be
V_{Ap}V^{A}{}_{q}=\eta_{pq}\,.
\label{Vdef}
\ee 
Consequently,  it generates a pair of symmetric and orthogonal projectors,
\be
\ba{lll}
\multicolumn{3}{c}{P_{AB}=P_{BA}:=V_{Ap}V_{B}{}^{p}\,,\quad\quad
\brP_{AB}=\brP_{BA}:=\cJ_{AB}-P_{AB}\,,}\\
P_{A}{}^{B}P_{B}{}^{C}=P_{A}{}^{C}\,,\quad&\quad
\brP_{A}{}^{B}\brP_{B}{}^{C}=\brP_{A}{}^{C}\,,\quad&\quad
P_{A}{}^{B}\brP_{B}{}^{C}=0\,.
\ea
\label{PbrP}
\ee
The DFT-vielbein may  convert the $\Off$ indices to $\Spinf$ vector indices and \textit{vice versa}, if there is no room for ambiguity, \textit{e.g.~}$\partial_{p}=V^{A}{}_{p}\partial_{A}$,   $\,\gamma^{A}=V^{A}{}_{p}\gamma^{p}$,  but $\,V_{A}{}^{p}\partial_{p}=P_{A}{}^{B}\partial_{B}\neq\partial_{A}$.

The higher spin gauge potential is anti-Hermitian, $\cW_{A}=-(\cW_{A})^{\dagger}$, and  the Majorana property of the bosonic spinor, $\Psi^{\alpha}$, means that its Dirac conjugate equals the charge conjugate, $\bar{\Psi}=\Psi^{\dagger}\mathbf{A}=\Psi^{T}\mathbf{C}$.

 It is crucial for us to postulate that -- as the arguments indicate --    $d(x)$ and $V_{Ap}(x)$ are independent of the internal  spinorial coordinates, and hence they are HS singlets, \textit{c.f.~}(\ref{HSGSsgl}). On the other hand, $\cW_{A}(x,\zeta,\brzeta)$ and $\Psi^{\alpha}(x,\zeta,\brzeta)$  are HS algebra valued: they depend  on  the internal  coordinates and thus transform  under the HS gauge symmetry in a nontrivial manner, \textit{c.f.~}(\ref{HSGS}).\\

\item \textit{Symmetry transformation rules.}
\begin{itemize}
\item The $\Off$ T-duality is a global symmetry rotating  the uppercase alphabet  indices of not only    the DFT-vielbein, $V_{Ap}$, and   the HS gauge potential,  $\cW_{A}$, but also   the coordinates themselves, $x^{A}$,    which are the common arguments in all the  fields in \eqref{FC}: 
with $L\in\Off$,  the finite transformations are given by
\be
\ba{lll}
\left(
\ba{c}
x^{A}\\
d(x)\\
V_{Ap}(x)\\
\cW_{A}(x,\zeta,\brzeta)\\
\Psi^{\alpha}(x,\zeta,\brzeta)
\ea
\right)
\quad&\longrightarrow&\quad
\left(
\ba{c}
x^{\prime A}=x^{B}L_{B}{}^{A}\,,\\
d(x^{\prime})\\
L_{A}{}^{B}V_{Bp}(x^{\prime})\\
L_{A}{}^{B}\cW_{B}(x^{\prime},\zeta,\brzeta)\\
\Psi^{\alpha}(x^{\prime},\zeta,\brzeta)
\ea
\right)\,.
\ea
\label{Offtransf}
\ee

\item The diffeomorphisms on the doubled-yet-gauged coordinate system,  \textit{i.e.~DFT-diffeomorphisms}, are generated  by the generalized Lie derivative~\cite{Siegel:1993th,Grana:2008yw}, such that 
\be
\ba{ll}
\delta_{X}d=X^{A}\partial_{A}d-\half\partial_{A}X^{A} \,,\quad&\quad
\delta_{X}V_{Ap}=X^{B}\partial_{B}V_{Ap}+(\partial_{A}X^{B}-\partial^{B}X_{A})V_{Bp}\,,\\
\delta_{X}\Psi^{\alpha}=X^{A}\partial_{A}\Psi^{\alpha}\,,\quad&\quad
\delta_{X}\cW_{A}=X^{B}\partial_{B}\cW_{A}+(\partial_{A}X^{B}-\partial^{B}X_{A})\cW_{B}\,,
\ea
\label{GLieDeriv}
\ee
where $X^{A}(x)$ is an $\Off$ vectorial diffeomorphism parameter which should be  independent of the internal spinorial  coordinates, $\zeta, \brzeta$.

\item The  higher spin gauge symmetry, $\HSf$, is realized through the adjoint action of the star product,
\be
\ba{ll}
\delta_{\cT}\Psi=[\cT,\Psi]_{\star}\,,\quad&\quad
\delta_{\cT}\cW_{A}=-\big(\partial_{A}\cT+[\cW_{A},\cT]_{\star}\big)\equiv-\cD_{A}\cT\,,
\ea
\label{HSGS}
\ee
where $\cT(x,\zeta,\brzeta)$ is a  local parameter of the HS  gauge symmetry,   which is anti-Hermitian, $\cT=-(\cT)^{\dagger}$, and   depends arbitrarily  on all the coordinates.  The DFT-dilaton and the DFT-vielbein are   independent of the internal spinorial coordinates, and hence they are higher spin gauge   singlets,
\be
\ba{ll}
\delta_{\cT}d=[\cT,d(x)]_{\star}=0\,,\quad&\quad \delta_{\cT}V_{Ap}
=[\cT,V_{Ap}(x)]_{\star}=0\,.
\ea
\label{HSGSsgl}
\ee

\item The $\Spinf$ local Lorentz symmetry  rotates  the explicit  unbarred lowercase indices, specifically   the  vectorial  small  roman  letter of $V_{Ap}$   and the spinorial greek letter of $\Psi^{\alpha}$\,: with a skew-symmetric, arbitrarily  $x$-dependent  local parameter, $\omega_{pq}(x)=-\omega_{qp}(x)$, its infinitesimal  transformations are  given by 
\be
\ba{ll}
\delta_{\omega}d(x)=0\,,\quad&\quad  
\delta_{\omega}V_{Ap}(x)=\omega_{p}{}^{q}(x)V_{Aq}(x)\,,  \\
\delta_{\omega}\cW_{A}(x,\zeta,\brzeta)=0\,,\quad&\quad
\delta_{\omega}\Psi^{\alpha}(x,\zeta,\brzeta)=\quarter\omega_{pq}(x)\left(\gamma^{pq}\right)^{\alpha}{}_{\beta}\Psi^{\beta}(x,\zeta,\brzeta)\,,
\ea
\label{Spinftransf}
\ee
where \textit{a priori}  the  internal spinorial  coordinates,   $\zeta^{\alpha}$, $\brzeta_{\beta}$, inside $\cW_{A}(x,\zeta,\brzeta)$ and $\Psi^{\alpha}(x,\zeta,\brzeta)$ are  {\bf{not}} rotated. However, the above   local Lorentz  transformations can be  modified to include  the  higher spin gauge symmetry with a particular form of  the local parameter,
\be
\cT_{\omega}(x,z)\equiv\textstyle{\frac{1}{4}}\omega_{pq}(x)\brzeta\gamma^{pq}\zeta\,.
\ee
\end{itemize}
From
\be
\ba{ll}
{}[\cT_{\omega},\zeta^{\alpha}]_{\star}=-\quarter\omega_{pq}(x)(\gamma^{pq}\zeta)^{\alpha}\,,\quad&\quad
{}[\cT_{\omega},\brzeta_{\alpha}]_{\star}=+\quarter\omega_{pq}(x)(\brzeta\gamma^{pq})_{\alpha}\,,
\ea
\label{ztransf}
\ee
the  modified $\Spinf$ local Lorentz transformation rule, $\delta_{\omega}\rightarrow\delta_{\omega+\cT_{\omega}}$,  then  rotates not only the explicit $\Spinf$ indices of the fields but also  all the internal spinorial  coordinates:  while  the HS gauge singlet fields transform in the same way  as before~\eqref{Spinftransf},
\be
\ba{ll}
\delta_{\omega+\cT_{\omega}}d=0\,,\qquad&\qquad
\delta_{\omega+\cT_{\omega}}V_{Ap}=\omega_{p}{}^{q}V_{Aq}\,, 
\ea
\label{modifiedSpinf1}
\ee
the HS gauge adjoint fields transform in a  novel way compared to (\ref{Spinftransf}),
\be
\ba{ll}
\delta_{\omega+\cT_{\omega}}\cW_{A}&
=-\partial_{A}\cT_{\omega}+[\cT_{\omega},\cW_{A}]_{\star}\\
{}&=-\textstyle{\frac{1}{4}}\partial_{A}\omega_{pq}\,\brzeta\gamma^{pq}\zeta
-\quarter\omega_{pq}(\gamma^{pq}\zeta)^{\beta}\partial_{\beta}\cW_{A}
+\quarter\omega_{pq}(\brzeta\gamma^{pq})_{\beta}\bar{\partial}^{\beta}\cW_{A}\,,\\
\delta_{\omega+\cT_{\omega}}\Psi^{\alpha}&=\quarter\omega_{pq}\left(\gamma^{pq}\right)^{\alpha}{}_{\beta}\Psi^{\beta}+[\cT_{\omega},\Psi]_{\star}\\
{}&=
\quarter\omega_{pq}\left(\gamma^{pq}\right)^{\alpha}{}_{\beta}\Psi^{\beta}
-\quarter\omega_{pq}(\gamma^{pq}\zeta)^{\beta}\partial_{\beta}\Psi^{\alpha}
+\quarter\omega_{pq}(\brzeta\gamma^{pq})_{\beta}\bar{\partial}^{\beta}\Psi^{\alpha}
\,,
\ea
\label{modifiedSpinf2}
\ee
where we set  $\partial_{\beta}\equiv\frac{\partial~}{\partial\zeta^{\beta}}$ and  $\bar{\partial}^{\beta}\equiv\frac{\partial~}{\partial\brzeta_{\beta}}$, and clearly all the spinors are to be rotated.   

In either case of \eqref{Spinftransf} or \eqref{modifiedSpinf2}, since the higher spin gauge symmetry does not rotate any of the explicit, external   indices of $V_{Ap}$ and $\Psi^{\alpha}$~(\ref{HSGS}), (\ref{HSGSsgl}), the  local Lorentz symmetry,  T-duality and the DFT-diffeomorphisms are all    intrinsically  different   from the higher spin gauge symmetry.\\

\item We define  \textit{the master  derivative} of HS-DFT,
\be
\cD_{A}=\partial_{A}+\Gamma_{A}(x)+\Phi_{A}(x)+\cW_{A}(x,\zeta,\brzeta)\,,
\label{Master}
\ee
which takes care of the DFT-diffeomorphisms~(\ref{GLieDeriv}), the  $\Spinf$ local Lorentz symmetry~(\ref{Spinftransf}), and the $\HSf$ gauge symmetry~(\ref{HSGS}), by employing the   relevant three connections:  

\begin{itemize}
\item \textit{(i)} the DFT   extension of the Christoffel connection~\cite{Jeon:2011cn}~(\textit{c.f.~}\cite{Jeon:2010rw}) ,
\be
\ba{ll}
\Gamma_{CAB}:=&2\left(P\partial_{C}P\brP\right)_{[AB]}
+2\left({{\brP}_{[A}{}^{D}{\brP}_{B]}{}^{E}}-{P_{[A}{}^{D}P_{B]}{}^{E}}\right)\partial_{D}P_{EC}\\
{}&-\textstyle{\frac{4}{3}}\left(\brP_{C[A}\brP_{B]}{}^{D}+P_{C[A}P_{B]}{}^{D}\right)\!\left(\partial_{D}d+(P\partial^{E}P\brP)_{[ED]}\right)\,,
\ea
\label{Gammao}
\ee

\item \textit{(ii)} the local Lorentz  spin connection~\cite{Jeon:2011sq,Jeon:2012hp},
\be
\Phi_{Apq}:=V^{B}{}_{p}(\partial_{A}V_{Bq}+\Gamma_{AB}{}^{C}V_{Cq})=V^{B}{}_{p}\na_{A}V_{Bq}\,,
\label{Phio}
\ee

\item \textit{(iii)} the higher spin gauge potential, $\cW_{A}$,  in the adjoint representation of the star product.  
\end{itemize}

In (\ref{Phio}), $\na_{A}$ denotes   the `semi-covariant derivative' set by the    DFT-Christoffel connection~\cite{Jeon:2011cn},
\be
\na_{A}:=\partial_{A}+\Gamma_{A}\,,
\ee
which satisfies, among others  (\textit{c.f.~}section~\ref{SECmaster}),
\be
\ba{llll}
\na_{A}\cJ_{BC}=0\,,\quad&\quad
\na_{A}P_{BC}=0\,,\quad&\quad
\na_{A}\brP_{BC}=0\,,\quad&\quad
\na_{A}d=0\,.
\ea
\ee

As an example,  note the expression,
\be
\cD_{A}\Psi^{\alpha}=\partial_{A}\Psi^{\alpha}
+\quarter\Phi_{Apq}(\gamma^{pq})^{\alpha}{}_{\beta}\Psi^{\beta}+\left[\cW_{A},\Psi^{\alpha}\right]_{\star}\,,
\ee
and also  the fact that the expression~(\ref{Phio}) follows from the compatibility of the master derivative with the DFT-vielbein which is HS singlet, 
\be
\cD_{A}V_{Bp}=\partial_{A}V_{Bp}+\Gamma_{AB}{}^{C}V_{Cp}+\Phi_{Ap}{}^{q}V_{Bq}=\na_{A}V_{Bp}+\Phi_{Ap}{}^{q}V_{Bq}=0\,.
\ee
By appropriate contractions with the projectors~(\ref{PbrP}) or the DFT-vielbein, the master  derivative can be completely covariantized, \textit{e.g.}~(\ref{HSDFTeq0}) -- (\ref{HSDFTeq3}). \\

\end{itemize}
\newpage


\subsection{Dynamics of HS-DFT}

\begin{itemize}

\item  The proposed  functional, \textit{i.e.~action,} of HS-DFT is 
\be
\ba{ll}
\dis{\cS_{\HSDFT}=\int\rd^{4} x~~\cL_{\HSDFT}\,,}\quad&\quad
\cL_{\HSDFT}=\cL_{\DFT}+\cL_{\sHS}\,,
\ea
\label{THEACTION}
\ee
where the  $x$-integral is to be taken over a four-dimensional section of choice, and the HS-DFT Lagrangian consists of  two parts: 

\begin{itemize}
\item  \textit{(i)} the `pure' DFT Lagrangian, 
\be
\cL_{\DFT}=e^{-2d}\Big[\,(P^{AB}P^{CD}-\brP^{AB}\brP^{CD})S_{ACBD}-2\Lambda_{\DFT}\,\Big]\,,
\label{pureDFTL}
\ee
\item \textit{(ii)} the   `matter' HS Lagrangian,
\be
\cL_{\sHS}=\gHS^{-2}\,e^{-2d}\,\Tr\left[P^{AC}\brP^{BD}\cF^{\cW}_{AB}\star\cF^{\cW}_{CD}
+\bar{\Psi}\star\gammaf\gamma^{A}\cD_{A}\Psi-V_{\star}(\Psi)
\right]\,.
\label{HSL}
\ee
\end{itemize}
~\\
The `pure' DFT Lagrangian~(\ref{pureDFTL}) contains   the DFT version of the cosmological constant, $\Lambda_{\DFT}$, of which the value can be arbitrary,  and  the `semi-covariant' four-index curvature~\cite{Jeon:2011cn},
\be
S_{ABCD}:=\half(R_{ABCD}+R_{CDAB}-\Gamma^{E}{}_{AB}\Gamma_{ECD})\,,
\ee
with
\be
R_{CDAB}:=\partial_{A}\Gamma_{BCD}-\partial_{B}\Gamma_{ACD}+\Gamma_{AC}{}^{E}\Gamma_{BED}-\Gamma_{BC}{}^{E}\Gamma_{AED}\,.
\ee

In the  `matter' HS Lagrangian~(\ref{HSL}), $\gHS$ denotes  a coupling constant; $V_{\star}(\Psi)$ is a star-producted scalar potential of the bosonic spinor,  $\Psi$;  and the trace means   the  formal $(\zeta,\brzeta)$-integrals over the (real eight-dimensional) internal space,
\be\label{trace}
\Tr\big[\quad\cdot\quad\big]:=\int\rd^{4}\zeta\int\rd^{4}\brzeta~ \big[\quad\cdot\quad\big]\,.
\ee
We  also set  the `semi-covariant' field strength of  the higher spin gauge potential, following \cite{Jeon:2011kp,Choi:2015bga},
\be
\cF^{\cW}_{AB}:=\na_{A}\cW_{B}-\na_{B}\cW_{A}+[\cW_{A},\cW_{B}]_{\star}\,.
\ee
Since the Majorana spinor, $\Psi$, is bosonic while $\mathbf{C}\gammaf\gamma^{A}$ is skew-symmetric, the kinetic term of $\Psi$, \textit{i.e.~}the second term in (\ref{HSL}),  is not a total derivative.  On account of  the cosmological constant, $\Lambda_{\DFT}$,  without loss of generality, we assume the `absolute' minimum of the potential  to vanish, \textit{i.e.~}$\min[V_{\star}(\Psi)]=0$.

We emphasize that each term in the Lagrangian  is completely covariant, under  all the symmetries: $\Off$ T-duality, DFT-diffeomorphisms, $\Spinf$ local Lorentz symmetry and $\HSf$ higher spin gauge symmetry. 

\item  \textit{The Euler-Lagrange equations}  of the full    action~(\ref{THEACTION}) are as follows. 

 The equation of motion of the DFT-dilaton implies that the on-shell Lagrangian should vanish,
\be
\cL_{\HSDFT}=0\,.
\label{EOMd}
\ee

For the DFT-vielbein, with $S_{AB}:=S_{ACB}{}^{C}$, we have both
\be
\bar{\Psi}\star\gamma^{pq}\gammaf\gamma^{A}\cD_{A}\Psi=0\,,
\label{EOMV}
\ee
and, as a HS-DFT extension of the  Einstein  equation,
\be
\!\! P_{A}{}^{C}\brP_{B}{}^{D}\!\left(\!
S_{CD}\!+\!\half\gHS^{-2}\Tr\Big[\!
\left\{\cF^{\cW}\star(P{-\brP})\cF^{\cW}\right\}_{CD}\!+\!
\na_{E}
\left(\cF^{\cW}_{CD}\star\cW^{E}\right)
\!+\!\half\bar{\Psi}\star\gammaf\gamma_{C}\cD_{D}\Psi\Big]\right)\!=\!0\,.
\label{EOMV2}
\ee
The equation of motion of the 
 higher spin gauge potential is also twofold, as it implies   both
\be
\cD_{B}\big(P\cF^{\cW}\brP\big)^{BA}=0\,,
\label{EOMW1}
\ee
and 
\be
\cD_{B}\big(P\cF^{\cW}\brP\big)^{AB}-\half
\big[\,\Psi^{\alpha},\Psi^{\beta}\,\big]_{\star}(\mathbf{C}\gammaf\gamma^{A})_{\alpha\beta}=0\,.
\label{EOMW2}
\ee

Finally, for  the bosonic   Majorana spinor, we have a HS-DFT extension of the Dirac equation,
\be
\gamma^{A}\cD_{A}\Psi-\half\gammaf\mathbf{C}^{-1}\,\partial_{\Psi}\Tr\left[V_{\star}(\Psi)\right]=0\,,
\label{EOMPsi}
\ee
which actually implies \eqref{EOMV} provided the potential, $V_{\star}(\Psi)$, is $\Spinf$ singlet.

\item The  full set of the Euler-Lagrange equations~(\ref{EOMd})\,--\,(\ref{EOMPsi}), are automatically  fulfilled,   provided  the following `stronger' equations, or   \textit{BPS-like conditions}, hold:
\begin{eqnarray}
&&(P^{AB}P^{CD}-\brP^{AB}\brP^{CD})S_{ACBD}-2\Lambda_{\DFT}-\gHS^{-2}\Tr[V_{\star}(\Psi)]=0\,,\label{HSDFTeqLDFT}\\
&&P_{A}{}^{C}\brP_{B}{}^{D}S_{CD}=0\,,\label{HSDFTeq0}
\end{eqnarray}
and
\begin{eqnarray}
&&P_{A}{}^{C}\brP_{B}{}^{D}\cF^{\cW}_{CD}=0\,,\label{HSDFTeq1}\\
&&\brP_{A}{}^{B}\cD_{B}\Psi=0\,,\label{HSDFTeq2}\\
&&{}\gamma^{A}\cD_{A}\Psi=0\,,\label{HSDFTeq3}\\
&&{}\left[\Psi^{\alpha},\Psi^{\beta}\right]_{\star}(\mathbf{C}\gammaf\gamma^{p})_{\alpha\beta}=0\,,
\label{HSDFTeq4}\\
&&\partial_{\Psi}\Tr[V_{\star}(\Psi)]=0\,.\label{HSDFTeq5}
\end{eqnarray}  
Especially, when $\gHS^{-2}\Tr[V_{\star}(\Psi)]$ vanishes (either in the large $\gHS$ limit or at the bottom of the potential),  (\ref{HSDFTeqLDFT}) and (\ref{HSDFTeq0}) become precisely the equations of motion of the `pure' DFT Lagrangian~(\ref{pureDFTL}),  and hence any solution to the above  BPS equations is qualified as a 	`vacuum' configuration.

The  `algebraic' condition~(\ref{HSDFTeq4}) is equivalent, from (\ref{complete2}), to
\be
\big[\Psi^{\alpha},\Psi^{\beta}\big]_{\star}=\quarter
\big[(1+\gammaf)\mathbf{C}^{-1}\big]^{\alpha\beta}\cQ_{+}
+\quarter\big[(1-\gammaf)
\mathbf{C}^{-1}\big]^{\alpha\beta}\cQ_{-}\,,
\label{PPQQ}
\ee
where we set
\be
\ba{l}
\cQ_{+}:=
-\bar{\Psi}\star(1+\gammaf)\star\Psi=
-\half\big[\Psi^{\alpha},\Psi^{\beta}\big]_{\star}\big[\mathbf{C}(1+\gammaf)\big]_{\alpha\beta}\,,\\
\cQ_{-}:=-
\bar{\Psi}\star(1-\gammaf)\star\Psi=
-\half\big[\Psi^{\alpha},\Psi^{\beta}\big]_{\star}\big[\mathbf{C}(1-\gammaf)\big]_{\alpha\beta}=(\cQ_{+})^{\dagger}\,.
\ea
\label{cQdef}
\ee
~\\

\item The HS-DFT  BPS equations, (\ref{HSDFTeqLDFT})\,--\,(\ref{HSDFTeq5}), and consequently the full set of the equations of motion, (\ref{EOMd})\,--\,(\ref{EOMPsi}), admit the following \textit{vacuum solution}, characterized by   linear DFT-dilaton and  constant, \textit{i.e.~}flat, DFT-vielbein  (\textit{c.f.~}section~\ref{SEClineard}),
\be
\ba{llll}
\mr{d}=\mr{N}_{A}x^{A}\,,
\qquad&\qquad\partial_{A}\mr{V}_{Bp}=0\,,\quad&\quad
\mr{\cW}_{A}=-\textstyle{\frac{1}{3}}\brzeta\gamma_{AB}\zeta\mr{N}^{B}\,,
\quad&\quad
\mr{\Psi}{}^{\alpha}= m^{\frac{3}{2}}\,\mathbf{Re}(\zeta^{\alpha})\,,
\ea
\label{lineardvacuum}
\ee
provided the potential assumes its minimum or $\gHS^{-2}\Tr[V_{\star}(\mr{\Psi})]=0$. 
Here $\mr{N}_{A}=\partial_{A}\mr{d}\,$ is  a  constant $\Off$ null vector, $\mr{N}_{A}\mr{N}^{A}=0$,  normalized  to meet
\be
\Lambda_{\DFT}=-4\mr{N}_{p}\mr{N}^{p}\,.
\ee
Further, $m$ is a constant mass parameter introduced to match  the $\frac{3}{2}$ mass dimension of $\Psi$, and $\mathbf{Re}(\zeta^{\alpha})$ denotes the `real' part of  $\zeta^{\alpha}$,
\be
\mathbf{Re}(\zeta^{\alpha})=
\half(\zeta^{\alpha}+\brzeta_{\beta}\mathbf{C}^{-1\beta\alpha}
)\,.
\ee
In particular, the vacuum solution gives
\be
\mr{\cQ}_{+}=\mr{\cQ}_{-}=m^{3}\,.
\ee
We emphasize that, in our proposal of HS-DFT, there is no restriction on  $\Lambda_{\DFT}\,$: any value with any sign is allowed, as is the case with half-maximal  supersymmetric gauged  double field theories~\cite{Cho:2015lha}. Accordingly, the above linear DFT-dilaton vacuum  calls  for a space-like, null-like or time-like four-dimensional vector, $\mr{N}^{p}$, in each case of  $\Lambda_{\DFT}{<0}$, $\Lambda_{\DFT}{=0}$ or  $\Lambda_{\DFT}{>0}$  respectively.\\

\item On the other hand,  imposing   extra   conditions: \textit{(i)} the sectioning condition on the HS gauge potential,
\be
\ba{ll}
\cW^{A}\partial_{A}\equiv0\,,\qquad&\quad
\cW^{A}\cW_{A}\equiv0\,,
\ea
\label{Wseccon} 
\ee
\textit{(ii)} the deformed  oscillator relations  on the bosonic Majorana spinor, as solutions to  \eqref{HSDFTeq5},
\be
\ba{ll}
\left\{(1{+\gammaf})\Psi\,,\,\cQ_{+}\right\}_{\star}\equiv 2m^{3}(1{+\gammaf})\Psi\,,\quad&\quad
\left\{(1{-\gammaf})\Psi\,,\,\cQ_{-}\right\}_{\star}
\equiv2m^{3}(1{-\gammaf})\Psi\,,
\ea
\label{PsicQ}
\ee
and \textit{(iii)} a twisted reality condition on $\cQ_{\pm}$ in terms of the inner Klein operators defined in \eqref{innerKlein},
\be
\cQ_{+}-m^{3}=\left(\cQ_{-}-m^{3}\right)\star\bfk\star\brbfk\,,
\label{Qreality}
\ee
we may reduce  the HS-DFT BPS conditions, (\ref{HSDFTeqLDFT})\,--\,(\ref{HSDFTeq5}),   to the bosonic   Vasiliev equations in four dimensions, \textit{c.f.~}(\ref{dftV1})\,--\,(\ref{dftV4}), \eqref{Raction}, \eqref{RVeq}. 
 The above linear dilaton vacuum~(\ref{lineardvacuum}) does not meet  (\ref{Wseccon}). Moreover, unlike the  Vasiliev theory,   its HS gauge potential, $\mr{\cW}_{A}$,  does not  contain a gravitational spin-two `vielbein' (only the spin connection appears, \textit{c.f.~}\eqref{mrPhi}).  Hence,  it corresponds to a genuine HS-DFT background, particularly   realizing the open string interpretation~\cite{Sagnotti:2013bha}. 
 
 \item In particular, the following  two  choices of the potential are of interest,
 \be
 V^{\YM}_{\star}(\Psi)=\half\lambda_{\YM}\big[\, (\cQ_{+}-m^{3})\star(\cQ_{+}-m^{3})\,+\,(\cQ_{-}-m^{3})\star(\cQ_{-}-m^{3})\,\big]\,,
 \label{VYM}
 \ee
 and
 \be
 V^{\Vasiliev}_{\star}(\Psi)=\half\lambda_{\Vasiliev}\big(\,
 \cR_{+}\star\cR_{+}\,+\,\cR_{-}\star\cR_{-}\,\big)\,,
 \label{VVasiliev}
 \ee
where for the latter  we set, similarly  to (\ref{cQdef}), 
 \be
\ba{ll}
{\cR}_{\pm}:=-\bar{\Upsilon}_{\pm}\star(1\pm\gammaf)\star\Upsilon_{\pm}\,,\quad&\quad
\Upsilon_{\pm}:=\left\{(1\pm\gammaf)\Psi\,,\,({\cQ_{\pm}-m^{3}})\right\}_{\star}\,.
\ea
\ee
We call the former  ``Yang-Mills'' potential and the latter  ``deformed oscillator'' potential, as the former is essentially  $\Psi$-commutator squared, \textit{c.f.~}(\ref{cQdef}), up to  surface integral over the internal space and constant shift, and the latter is designed to make the  deformed  oscillator relations~(\ref{PsicQ}), \textit{i.e.}~$\Upsilon_{+}=\Upsilon_{-}=0$, to solve    the  algebraic  BPS condition~\eqref{HSDFTeq5}, \textit{i.e.~}$\partial_{\Psi}\Tr[V^{\Vasiliev}_{\star}(\Psi)]{=0}$.

In the case of  the ``Yang-Mills'' potential,   we have  $\cQ_{\pm}\rightarrow m^{3}$ in the low  energy  limit, and hence the deformed oscillator relations can be approximately achieved, while it may realize a Brout-Englert-Higgs mechanism. It is worth while to note for the last algebraic BPS condition~(\ref{HSDFTeq5}),
\be
\ba{lll}
\partial_{\Psi}\Tr\left[V^{\YM}_{\star}(\Psi)\right]=0\quad&\Longleftrightarrow&\quad
\left[(1{+\gammaf})\Psi\,,\,\cQ_{+}\right]=0\,,~~~\&~~~
\left[(1{-\gammaf})\Psi\,,\,\cQ_{-}\right]=0\,.
\ea
\label{YMBPS}
\ee

In summary, the HS-DFT extension of the bosonic Vasiliev equations consists of \eqref{HSDFTeq1}, \eqref{HSDFTeq2}, \eqref{HSDFTeq3}, \eqref{HSDFTeq4}, \eqref{Wseccon} and \eqref{PsicQ}; while the BPS conditions~\eqref{HSDFTeq1}\,--\,\eqref{HSDFTeq5},  together with the ``YM'' potential~(\ref{VYM}), may lead to a  possible open string realization of  higher spin theory.

\end{itemize}

\newpage


\section{Exposition \label{SECEXPOSITION}}

Here  we provide some complementary  explanations  of  the HS-DFT  proposed in the previous  section. 

\subsection{Doubled-yet-gauged  coordinates   and DFT-diffeomorphisms\label{SECDyG}}

The section condition~\eqref{seccon} decomposes into the linear weak  constraint and the  quadratic strong constraint:
\be
\ba{ll}
\partial_{A}\partial^{A}\phi= 0\,,\quad&\quad
\partial_{A}\phi\,\partial^{A}\varphi= 0\,.
\ea
\ee 
Here and in this subsection, $\phi$ and $\varphi$ are arbitrary  functions and their derivatives in the HS-DFT we construct.  Demanding the weak constraint to hold also for a product, \textit{i.e.~}$\partial_{A}\partial^{A}(\phi\varphi)=0$, we are led to   the strong constraint.  This  explains the nomenclature    behind the terms,  `weak' and `strong'.  On the other hand, if we substitute $\partial^{B}\phi$ and $\partial_{C}\phi$ into $\phi$ and $\varphi$, the strong constraint actually implies the weak constraint, since  as an   $8\times 8$ matrix,   \,$\partial_{A}\partial^{B}\phi$\, is  nilpotent and hence is  traceless~\cite{Lee:2013hma}.  Furthermore, as can be shown easily from  the  power series expansion~\cite{Park:2013mpa}, the strong constraint means that all the functions in the theory are invariant under  the following   `shifts' of the doubled coordinates,
\be
\ba{ll}
\Phi(x+\Delta)=\Phi(x)\,,\quad&\quad\Delta^{A}=\phi\,\partial^{A}\varphi\,.
\ea
\label{aTensorCGS}
\ee
This simple observation  reveals the geometric meaning behind  the section condition that, \textit{the doubled coordinates  are in fact  gauged}  through  an  equivalence relation~\cite{Park:2013mpa,Lee:2013hma},
\be
x^{A}~\sim~x^{A}+\phi\,\partial^{A}\varphi\,.
\label{aCGS}
\ee 
It is then not a point in the doubled coordinate system but   
each equivalence class or a gauge orbit that  represents a single physical point in the undoubled spacetime (\textit{c.f.~}\cite{Berman:2014jba,Hull:2014mxa,Naseer:2015tia,Rey:2015mba}
 for  further discussions).\footnote{This  idea of `Coordinate Gauge Symmetry' can be  realized on a string worldsheet  as a conventional gauge symmetry of a doubled sigma model, by introducing a corresponding gauge potential~\cite{Lee:2013hma}. Integrating out the auxiliary gauge potential, the doubled sigma model reduces to the standard undoubled string  action   on an arbitrarily  curved NS-NS background. } \\

With the decomposition of the doubled coordinates, $x^{A}=(\tx_{\mu},x^{\nu})$, with respect to the  $\Off$ invariant metric, $\cJ_{AB}$~(\ref{cJ}), and  from $\partial_{A}\partial^{A}=2\frac{\partial^{2}~}{\partial\tx_{\mu}\partial x^{\mu}}$,  the section condition can be conveniently solved by requiring that all the fields are independent of    the ``dual'' coordinates,
\be
\frac{\partial~~}{\partial\tx_{\mu}}\equiv0\,.
\label{conventionalSEC}
\ee
The general solutions are then given  by the  $\Off$ rotations of this specific  solution. However, unless mentioned explicitly, we shall not assume any particular solution to the section condition such as \eqref{conventionalSEC}, as we intend to keep the manifest $\Off$ covariance throughout the proposal. \\

The {diffeomorphisms} on the doubled-yet-gauged coordinate system,  \textit{i.e.~DFT-diffeomorphisms},   are generated by the  generalized Lie derivative~\cite{Siegel:1993th,Grana:2008yw}. Acting on an arbitrary covariant tensor with weight $\omega_{{\scriptscriptstyle{T\,}}}$, it reads  
\be
\hcL_{X}T_{A_{1}\cdots A_{n}}:=X^{B}\partial_{B}T_{A_{1}\cdots A_{n}}+\omega_{{\scriptscriptstyle{T\,}}}\partial_{B}X^{B}T_{A_{1}\cdots A_{n}}+\sum_{i=1}^{n}(\partial_{A_{i}}X_{B}-\partial_{B}X_{A_{i}})T_{A_{1}\cdots A_{i-1}}{}^{B}{}_{A_{i+1}\cdots  A_{n}}\,,
\label{tcL}
\ee
where other types of indices are suppressed for simplicity. 
Specifically, for the  field content of HS-DFT~(\ref{FC}), the DFT-diffeomorphisms are given by
\be
\ba{l}
\delta_{X}d=-\half e^{2d}\hcL_{X}\left(e^{-2d}\right)=X^{A}\partial_{A}d-\half\partial_{A}X^{A}\,,\\
\delta_{X}V_{Ap}=\hcL_{X}V_{Ap}=X^{B}\partial_{B}V_{Ap}
+(\partial_{A}X^{B}-\partial^{B}X_{A})V_{Bp}\,,\\
\delta_{X}\cW_{A}=\hcL_{X}\cW_{A}=X^{B}\partial_{B}\cW_{A}
+(\partial_{A}X^{B}-\partial^{B}X_{A})\cW_{B}\,,\\
\delta_{X}\Psi^{\alpha}=\hcL_{X}\Psi^{\alpha}=X^{A}\partial_{A}\Psi^{\alpha}\,.
\ea
\ee
It is worth while to note that the $\Off$ invariant metric is compatible with  the generalized Lie derivative,  $\hcL_{X}\cJ_{AB}=0$, and 
the commutator of the generalized Lie derivatives is closed, up to the section condition,   by so-called the  C-bracket~\cite{Siegel:1993th,Hull:2009zb},
\be
\ba{ll}
\left[\hcL_{X},\hcL_{Y}\right]=\hcL_{[X,Y]_{\rmC}}\,,\quad&\quad
[X,Y]^{A}_{\rmC}= X^{B}\partial_{B}Y^{A}-Y^{B}\partial_{B}X^{A}+\half Y^{B}\partial^{A}X_{B}-\half X^{B}\partial^{A}Y_{B}\,.
\ea
\label{Cbracket}
\ee

\subsection{$\Spinf$ Clifford algebra: Majorana and Dirac spinors}
Combining $\mathbf{A}$ and $\mathbf{C}$, (\ref{mathbfA}), (\ref{mathbfC}),   we obtain
\be
\ba{lll}
\mathbf{B}_{\alpha\beta}=\mathbf{C}_{\alpha\gamma}\mathbf{A}^{\gamma}{}_{\beta}\,,\quad&\quad
(\gamma^{p})^{\ast}=+\mathbf{B}\gamma^{p}\mathbf{B}^{-1}\,,~~~~&~~~~ \left(\gamma^{(5)}\right)^{\ast}=-\mathbf{B}\gamma^{(5)}\mathbf{B}^{-1}\,.
\ea
\label{BM}
\ee
Consequently,    with  $\mathbf{A}=\mathbf{A}^{\dagger}=\mathbf{C}^{-1}\mathbf{B}$, we get
\be
\ba{ll}
(\mathbf{C}^{-1})^{\dagger}=+\mathbf{B}\mathbf{C}^{-1}\mathbf{B}\,,\quad&\quad
(\gammaf \mathbf{C}^{-1})^{\dagger}=-\mathbf{B}\gammaf \mathbf{C}^{-1}\mathbf{B}\,.
\ea
\label{CgammafC}
\ee
It is crucial to note that  $\mathbf{B}$ is unitary and symmetric, satisfying~\cite{Kugo:1982bn},
\be
\ba{llll}
\mathbf{B}^{\ast}\mathbf{B}=1\,,~~~~&~~~~ \mathbf{B}_{\alpha\beta}=\mathbf{B}_{\beta\alpha}\,.
\ea
\label{BBBB}
\ee
This property enables us to   impose  the Majorana conditions  on the spinor, $\Psi^{\alpha}$,  
\be
\ba{lll}
\bar{\Psi}=\Psi^{\dagger}\mathbf{A}=\Psi^{T}\mathbf{C}&\quad
\Longleftrightarrow\quad&
\Psi^{\ast}=\mathbf{B}\Psi\,,
\ea
\label{Majorana}
\ee
in a self-consistent manner, as
\be
\Psi=(\Psi^{\ast})^{\ast}=(\mathbf{B}\Psi)^{\ast}
=\mathbf{B}^{\ast}\mathbf{B}\Psi\,.
\ee
Further, we can decompose  the complex  Dirac spinor, $\zeta^{\alpha}$, which sets the non-commutative internal space~(\ref{zetacom}),  into  the  `real' and  `imaginary'  parts,
\be
\zeta=\mathbf{Re}(\zeta)+\mathbf{Im}(\zeta)\,,
\ee
where
\be
\ba{lll}
\mathbf{Re}(\zeta)=\half
\left[\zeta+(\mathbf{B}\zeta)^{\ast}\right]
\equiv\zeta_{+}\quad
&\Longleftrightarrow&\quad\zeta_{+}^{\alpha}
=\half(\zeta^{\alpha}+\brzeta_{\beta}C^{-1\beta\alpha})\,,\\
\mathbf{Im}(\zeta)=\half
\left[\zeta-(\mathbf{B}\zeta)^{\ast}\right]\equiv\zeta_{-}
\quad&\Longleftrightarrow&\quad
\zeta_{-}^{\alpha}=
\half(\zeta^{\alpha}-\brzeta_{\beta}C^{-1\beta\alpha})\,.
\ea
\label{ReIm}
\ee
These are `real' (Majorana) and `imaginary' (pseudo-Majorana)  in the following sense, 
\be
\ba{lll}
\left(\mathbf{B}\zeta_{+}\right)^{\ast}=+\zeta_{+}\,,~\quad&\quad
\left(\mathbf{B}\zeta_{-}\right)^{\ast}=-\zeta_{-}\,,~\quad&\quad
\brzeta_{\pm}=\zeta_{\pm}^{\dagger}\mathbf{A}=\pm\zeta_{\pm}^{T}\mathbf{C}\,.
\ea
\label{MpseudoM}
\ee
For the bosonic Dirac spinors, $\zeta$, $\brzeta$,  we have then\footnote{On the other hand, for a \textit{fermionic} (\textit{i.e.} odd Grassmannian) Dirac spinor, say $\psi$,  the relation~\eqref{barpsi2} gets modified,
\[\bar{\psi}\psi=\bar{\psi}_{+}\psi_{+}+\bar{\psi}_{-}\psi_{-}=
\psi_{+}^{T}\mathbf{C}\psi_{+}-\psi_{-}^{T}\mathbf{C}\psi_{-}\,.\]}
\be
\bar{\zeta}\zeta=(\bar{\zeta}\zeta)^{\dagger}
=(\bar{\zeta}\zeta)^{\ast}=\bar{\zeta}_{+}\zeta_{-}+\bar{\zeta}_{-}\zeta_{+}
=2\zeta_{+}^{T}\mathbf{C}\zeta_{-}=-2\zeta_{-}^{T}\mathbf{C}\zeta_{+}\,.
\label{barpsi2}
\ee
Our spacetime signature, $(-+++)$, admits the real, \textit{i.e.~}Majorana, representation of the gamma matrices, which means that, if desired,  we may put  $\,\mathbf{B}\equiv 1$, \textit{c.f.~}(\ref{ggAABBCC}). 

~\\
Both the higher spin gauge potential, $\cW_{A}$, and the bosonic  Majorana spinor, $\Psi^{\alpha}$, are HS algebra valued, such that they depend on   all the coordinates, $x^{A},\zeta^{\alpha},\brzeta_{\beta}$ generically, and can be expanded by the  internal spinorial   coordinates,\footnote{For a duality manifest  alternative approach to higher spin fields through twistor variables, see Ref.\cite{Cederwall:2015jfa} by Cederwall.} 
\be
\ba{l}
\dis{\cW_{A}(x,\zeta,\brzeta)=\sum_{m,n}\frac{1}{m!n!}\,\zeta^{\alpha_{1}}\zeta^{\alpha_{2}}
\cdots\zeta^{\alpha_{m}}\brzeta_{\beta_{1}}\brzeta_{\beta_{2}}
\cdots\brzeta_{\beta_{n}}\cW_{A\alpha_{1}\alpha_{2}\cdots
\alpha_{m}}{}^{\beta_{1}\beta_{2}\cdots\beta_{n}}(x)\,,}\\
\dis{\,\Psi^{\alpha}(x,\zeta,\brzeta)\,=\,\sum_{m,n}\frac{1}{m!n!}\,\zeta^{\alpha_{1}}\zeta^{\alpha_{2}}
\cdots\zeta^{\alpha_{m}}\brzeta_{\beta_{1}}\brzeta_{\beta_{2}}
\cdots\brzeta_{\beta_{n}}\Psi^{\alpha}{}_{\alpha_{1}\alpha_{2}\cdots
\alpha_{m}}{}^{\beta_{1}\beta_{2}\cdots\beta_{n}}(x)
\,.}
\ea
\label{expansion}
\ee
Naturally, the component fields  satisfy   symmetric properties,
\be
\ba{l}
\dis{\cW_{A\alpha_{1}\alpha_{2}\cdots
\alpha_{m}}{}^{\beta_{1}\beta_{2}\cdots\beta_{n}}(x)=
\cW_{A(\alpha_{1}\alpha_{2}\cdots
\alpha_{m})}{}^{(\beta_{1}\beta_{2}\cdots\beta_{n})}(x)\,,}\\
\dis{\Psi^{\alpha}{}_{\alpha_{1}\alpha_{2}\cdots
\alpha_{m}}{}^{\beta_{1}\beta_{2}\cdots\beta_{n}}(x)=\Psi^{\alpha}{}_{(\alpha_{1}\alpha_{2}\cdots
\alpha_{m})}{}^{(\beta_{1}\beta_{2}\cdots\beta_{n})}(x)\,,}
\ea
\label{comp}
\ee
and also,  from the anti-Hermiticity of $\cW_{A}$ and the Majorana property  of $\Psi^{\alpha}$,     a set of reality conditions follows
\be
\ba{cll}
\mathbf{A}^{\beta_{1}}{}_{\delta_{1}}\cdots\mathbf{A}^{\beta_{n}}{}_{\delta_{n}}\cW_{A\alpha_{1}\cdots\alpha_{m}}{}^{\delta_{1}\cdots\delta_{n}}&=&
-\left(\mathbf{A}^{\alpha_{1}}{}_{\gamma_{1}}\cdots\mathbf{A}^{\alpha_{m}}{}_{\gamma_{m}}\cW_{A\beta_{1}\cdots\beta_{n}}{}^{\gamma_{1}\cdots\gamma_{m}}\right)^{\ast}\,,\\
\mathbf{B}_{\alpha\beta}\mathbf{A}^{\beta_{1}}{}_{\delta_{1}}\cdots\mathbf{A}^{\beta_{n}}{}_{\delta_{n}}\Psi^{\beta}{}_{\alpha_{1}\cdots\alpha_{m}}{}^{\delta_{1}\cdots\delta_{n}}&=&
+\left(\mathbf{A}^{\alpha_{1}}{}_{\gamma_{1}}\cdots\mathbf{A}^{\alpha_{m}}{}_{\gamma_{m}}\Psi^{\alpha}{}_{\beta_{1}\cdots\beta_{n}}{}^{\gamma_{1}\cdots\gamma_{m}}\right)^{\ast}\,.
\ea
\ee
These reality conditions   relate  the component fields pairwise, \textit{i.e.~}$m\leftrightarrow n$. \\

Similarly,  the higher spin gauge  parameter~(\ref{HSGS}) can be expanded,
\be
\ba{l}
\dis{\cT(x,\zeta,\brzeta)\,=\,\sum_{m,n}\frac{1}{m!n!}\,\zeta^{\alpha_{1}}\zeta^{\alpha_{2}}
\cdots\zeta^{\alpha_{m}}\brzeta_{\beta_{1}}\brzeta_{\beta_{2}}
\cdots\brzeta_{\beta_{n}}\cT_{\alpha_{1}\alpha_{2}\cdots
\alpha_{m}}{}^{\beta_{1}\beta_{2}\cdots\beta_{n}}(x)\,,}\\
\mathbf{A}^{\beta_{1}}{}_{\delta_{1}}\cdots\mathbf{A}^{\beta_{n}}{}_{\delta_{n}}\cT_{\alpha_{1}\cdots\alpha_{m}}{}^{\delta_{1}\cdots\delta_{n}}\,=\,
-\left(\mathbf{A}^{\alpha_{1}}{}_{\gamma_{1}}\cdots\mathbf{A}^{\alpha_{m}}{}_{\gamma_{m}}\cT_{\beta_{1}\cdots\beta_{n}}{}^{\gamma_{1}\cdots\gamma_{m}}\right)^{\ast}\,.
\ea
\ee
And clearly,   the component fields of different ranks of $(m,n)$~(\ref{comp}) may  transform to each other under the higher spin gauge symmetry, (\ref{HSGS}).\\

\subsection{Wick  ordered star product}

We recall the definition of the star product~(\ref{starproduct}),
\be
f(x,\zeta,\brzeta)\star g(x,\zeta,\brzeta)=\dis{f(x,\zeta,\brzeta)~\exp\left(
\frac{\lpartial~~}{\partial{\zeta}^{\alpha}}
\frac{\rpartial~~}{\partial \brzeta_{\alpha}}\right)g(x,\zeta,\brzeta)}\,.
\label{star1}
\ee
From the Hermiticity of the matrix, $\mathbf{A}$,  we note
\be
\left(f\star g\right)^{\dagger}=g^{\dagger}\star f^{\dagger}\,.
\ee
The star product can be equivalently reformulated as integrals,    
\be
f(x,\zeta,\brzeta)\star g(x,\zeta,\brzeta)=\textstyle{\frac{1}{(2\pi)^{4}}}\dis{\int\rd^{4} \lambda_{+}\int\rd^{4}\rho_{+}~~e^{\bar{\lambda}_{+}\rho_{+}}f(x,\zeta+\lambda_{+},\brzeta)g(x,\zeta,\brzeta+\bar{\rho}_{+})\,,}
\label{star2}
\ee
where $\lambda_{+}$ and $\rho_{+}$ are two separate  bosonic Majorana spinors to integrate, 
\be
\ba{ll}
\bar{\lambda}_{+}=\lambda_{+}^{\dagger}\mathbf{A}=\lambda_{+}^{T}\mathbf{C}\,,\quad&\quad
\bar{\rho}_{+}=\rho_{+}^{\dagger}\mathbf{A}=\rho_{+}^{T}\mathbf{C}\,.
\ea
\ee
In particular,  the product,   $\bar{\lambda}_{+}\rho_{+}$,  is pure imaginary, 
\be
(\bar{\lambda}_{+}\rho_{+})^{\dagger}=\bar{\rho}_{+}\lambda_{+}=
-\bar{\lambda}_{+}\rho_{+}\,,
\ee
and  its exponentiation can serve as an integrand  for an integral representation  of  the Dirac delta function,
\be
\textstyle{\frac{1}{(2\pi)^{4}}}\dis{\int\rd^{4}\rho_{+}~e^{\bar{\lambda}_{+}\rho_{+}}=\delta(\lambda_{+})\,.}
\ee
The equivalence of the two expressions for the star product, (\ref{star1}) and (\ref{star2}), can be  then  straightforwardly established.\footnote{Yet, the integral formula~(\ref{star2}) may have better convergence property  than the differential one~(\ref{star1}).}  We present our own proof in the Appendix (\ref{Appendixequivalence}).\\

The star product satisfies the associativity, 
\be
f\star\left( g \star h\right)=\left(f\star g\right)\star h=f\star g \star h\,,
\label{associativity}
\ee
which can be also shown directly from the integral expression of the star product, \textit{c.f.~}Appendix (\ref{Appendixassociativity}).\\

Further, the star product over the $(\zeta^{\alpha},\brzeta_{\beta})$ internal space is  isomorphic to the Wick ordered operator formalism,
\be
\colon f(\hzeta,\hbrzeta)\colon\colon g(\hzeta,\hbrzeta)\colon\,=\,\hcO\!\left[f(\zeta,\brzeta)\star g(\zeta,\brzeta)\right]\,,
\label{isomorphism}
\ee
where any hatted object is an operator; the colon denotes the Wick ordering to place all the unbarred (annihilation) operators, 
 $\hzeta^{\alpha}$,  to the right and    the barred  (creation)  operators, $\hbrzeta_{\beta}$, to the left; and for an arbitrary function of the internal commuting coordinates, $f(\zeta,\brzeta)$,  the corresponding  operator, $\hcO[f(\zeta,\brzeta)]$, is defined  subject to  the Wick   ordering  prescription, 
\be
\hcO\!\left[f(\zeta,\brzeta)\right]=\,\colon f(\hzeta,\hbrzeta) \colon\,.
\ee
We refer readers to  Appendix (\ref{Appendixisomorphism}) for our own proof of the isomorphism.\\

We  define a pair of  inner Klein operators,  $\bfk$ and $\brbfk$,   
exponentiating  the  quadratic  forms of the bosonic  internal  spinors~\cite{Vasiliev:1992av}:
\be
\ba{ll}
\bfk:=e^{-\brzeta(1+\gammaf)\zeta}\,,\quad&\quad
\brbfk:=e^{-\brzeta(1-\gammaf)\zeta}=\bfk^{\dagger}\,.
\ea
\label{innerKlein}
\ee
For an arbitrary function, $f(x,\zeta,\brzeta)$, we compute to acquire  
\be
\ba{l}
\bfk\star f(x,\zeta,\brzeta)\star\bfk\\
=\left\{\bfk\star f(x,\zeta,\brzeta)\right\}\star\bfk\\
=\left\{e^{-\brzeta(1+\gammaf)\zeta}\exp\left(
\frac{\lpartial~~}{\partial{\zeta}^{\alpha}}
\frac{\rpartial~~}{\partial \brzeta_{\alpha}}\right)
f(x,\zeta,\brzeta)\right\}\exp\left(
\frac{\lpartial~~}{\partial{\zeta}^{\beta}}
\frac{\rpartial~~}{\partial \brzeta_{\beta}}\right)
e^{-\brzeta(1+\gammaf)\zeta}\\
=\left\{e^{-\brzeta(1+\gammaf)\zeta}\exp\left(
-\big[\brzeta(1+\gammaf)\big]_{\alpha}
\frac{\rpartial~~}{\partial \brzeta_{\alpha}}\right)
f(x,\zeta,\brzeta)\right\}\exp\left(
-\frac{\lpartial~~}{\partial{\zeta}^{\beta}}\big[(1+\gammaf)\zeta\big]^{\beta}
\right)
e^{-\brzeta(1+\gammaf)\zeta}\\
=\left\{e^{-\brzeta(1+\gammaf)\zeta}
f(x,\zeta,-\brzeta\gammaf)\right\}\exp\left(
-\frac{\lpartial~~}{\partial{\zeta}^{\beta}}\big[(1+\gammaf)\zeta\big]^{\beta}
\right)
e^{-\brzeta(1+\gammaf)\zeta}\\
=\left\{e^{\brzeta(\gammaf+1)\zeta}
f(x,-\gammaf\zeta,-\brzeta\gammaf)\right\}
e^{-\brzeta(1+\gammaf)\zeta}\\
=f(x,-\gammaf\zeta,-\brzeta\gammaf)\,.
\ea
\label{kk}
\ee
Similarly, by replacing $\gammaf$ by $-\gammaf$, we  get
\be
\brbfk\star f(x,\zeta,\brzeta)\star\brbfk=f(x,{\gammaf\zeta},{\brzeta\gammaf})\,.
\ee
Then,   considering  various  cases of $f(x,\zeta,\brzeta)$, such as   constant, or  the Klein operators  themselves, we can obtain   the crucial properties of the Klein operators:
\be
\ba{lll}
\bfk\star\bfk=1\,,\qquad&\qquad
\brbfk\star\brbfk=1\,,\qquad&\qquad
\bfk\star\brbfk=
\brbfk\star\bfk\,,
\ea
\ee
and
\be
\ba{ll}
\bfk\star f(x,\zeta,\brzeta)=f(x,-\gammaf\zeta,-\brzeta\gammaf)\star\bfk\,,\quad&\quad
\brbfk\star f(x,\zeta,\brzeta)=f(x,+\gammaf\zeta,+\brzeta\gammaf)\star\brbfk\,.
\ea
\ee
Further, combining the two Klein operators, we  get 
\be
\bfk\star \brbfk\star f(x,\zeta,\brzeta)=f(x,-\zeta,-\brzeta)\star\bfk\star\brbfk\,.
\ee
Explicitly, in a similar fashion to \eqref{kk}, we have
\be
\bfk\star \brbfk=\brbfk\star \bfk=e^{-2\brzeta\zeta}\,.
\ee
~\\

The `bosonic' truncation of the HS-DFT is then achieved by requiring
\be
\ba{lll}
\bfk\star \brbfk\star \cW_{A}-\cW_{A}\star \bfk\star \brbfk=0\,,\quad&~~~
\bfk\star \brbfk\star \Psi^{\alpha}+\Psi^{\alpha}\star \bfk\star \brbfk=0\,,\quad&~~~
\bfk\star \brbfk\star \cT-\cT\star \bfk\star \brbfk=0\,,
\ea
\ee
and hence, equivalently,
\be
\ba{lll}
\cW_{A}(x,-\zeta,-\brzeta)=+\cW_{A}(x,\zeta,\brzeta)\,,~&~
\Psi^{\alpha}(x,-\zeta,-\brzeta)=-\Psi^{\alpha}(x,\zeta,\brzeta)\,,~&~
\cT(x,-\zeta,-\brzeta)=+\cT(x,\zeta,\brzeta)\,.
\ea
\label{bTrunc}
\ee
Basically the truncation implies that the HS gauge potential, $\cW_{A}$, and the HS gauge parameter,  $\cT$, are restricted to be even functions of  $\zeta, \brzeta$, while the Majorana spinor, $\Psi^{\alpha}$, should be an odd function. \\

It is straightforward to verify 
\be
\ba{ll}
{}\left[\brzeta\gamma^{r}\zeta\,,\,\zeta^{\alpha}\right]_{\star}=-(\gamma^{r}\zeta)^{\alpha}\,,\quad&\quad
{}\left[\brzeta\gamma^{r}\zeta\,,\,
\brzeta_{\alpha}\right]_{\star}
=(\brzeta\gamma^{r})_{\alpha}\,,\\
{}\left[\brzeta\gamma^{pq}\zeta\,,\,\zeta^{\alpha}\right]_{\star}=-(\gamma^{pq}\zeta)^{\alpha}\,,\quad&\quad
{}\left[\brzeta\gamma^{pq}\zeta\,,\,\brzeta_{\alpha}\right]_{\star}
=(\brzeta\gamma^{pq})_{\alpha}\,,
\ea
\label{forPMy}
\ee
and 
\be
\ba{rll}
{}\left[\brzeta\gamma^{r}\zeta\,,\,\brzeta\gamma^{s}\zeta\right]_{\star}&=&
\brzeta(\gamma^{r}\gamma^{s}-\gamma^{s}\gamma^{r})
\zeta~=~2\brzeta\gamma^{rs}\zeta\,,\\
{}\left[\brzeta\gamma^{pq}\zeta\,,\,\brzeta\gamma^{r}\zeta\right]_{\star}&=&
\brzeta(\gamma^{pq}\gamma^{r}-\gamma^{r}\gamma^{pq})
\zeta~=~2\left(
\eta^{qr}\brzeta\gamma^{p}\zeta
-\eta^{pr}\brzeta\gamma^{q}\zeta\right)\,,\\
{}\left[\brzeta\gamma^{pq}\zeta\,,\,\brzeta\gamma^{rs}\zeta\right]_{\star}&=&
\brzeta(\gamma^{pq}\gamma^{rs}-\gamma^{rs}\gamma^{pq})
\zeta~=~2\left(
\eta^{qr}\brzeta\gamma^{ps}\zeta
-\eta^{pr}\brzeta\gamma^{qs}\zeta
+\eta^{ps}\brzeta\gamma^{qr}\zeta
-\eta^{qs}\brzeta\gamma^{pr}\zeta\right)\,.
\ea
\label{yyyycom}
\ee
Therefore, if we define two sets of  Wick ordered quadratic operators,
\be
\ba{ll}
\hP{}^{r}:=\textstyle{\frac{1}{2}}\hbrzeta\gamma^{r}\hzeta=-\big(\hP^{r\,}\big)^{\dagger}\,,\quad&\quad
\hM{}^{pq}:=\textstyle{\frac{1}{2}}\hbrzeta\gamma^{pq}\hzeta=-\big(\hM{}^{pq\,}\big)^{\dagger}\,,
\ea
\label{defPM}
\ee
Eq.(\ref{forPMy}) gives
\be
\ba{ll}
{}[\hP{}^{r}\,,\,\hzeta^{\alpha}]=-\half(\gamma^{r}\hzeta)^{\alpha}\,,\quad&\quad
{}[\hP{}^{r}\,,\,\hbrzeta_{\alpha}]=\half(\hbrzeta\gamma^{r})_{\alpha}\,,\\
{}[\hM{}^{pq}\,,\,\hzeta^{\alpha}]=-\half(\gamma^{pq}\hzeta)^{\alpha}\,,\quad&\quad
{}[\hM{}^{pq}\,,\,\hbrzeta_{\alpha}]=
\half(\hbrzeta\gamma^{pq})_{\alpha}\,,
\ea
\label{PMztransf}
\ee
and Eq.(\ref{yyyycom}) realizes an $\so(2,3)$   algebra,
\be
\ba{rll}
{}[\hP{}^{r}\,,\,\hP{}^{s}]&=&\hM{}^{rs}\,,\\
{}[\hM{}^{pq}\,,\,\hP{}^{r}]&=&\eta^{qr}\hP{}^{p}
-\eta^{pr}\hP{}^{q}\,,\\
{}[\hM{}^{pq}\,,\,\hM{}^{rs}]&=&
\eta^{qr}\hM{}^{ps}
-\eta^{pr}\hM{}^{qs}+
\eta^{ps}\hM{}^{qr}-\eta^{qs}\hM{}^{pr}\,.
\ea
\label{so23}
\ee
~\\

Finally, it is worth while to note the star  commutator relations for the real (Majorana) and the imaginary (pseudo-Majorana) spinors,
\be
\ba{lll}
\big[\,\zeta^{\alpha}_{+}\,,\,\zeta^{\beta}_{+}\,\big]_{\star}=+\half \mathbf{C}^{-1\alpha\beta}\,,\quad&\quad
\big[\,\zeta^{\alpha}_{-}\,,\,\zeta^{\beta}_{-}\,\big]_{\star}=-\half \mathbf{C}^{-1\alpha\beta}\,,\quad&\quad
\big[\,\zeta^{\alpha}_{+}\,,\,\zeta^{\beta}_{-}\,\big]_{\star}=0\,,
\ea
\label{zetacom2}
\ee
which are equivalent to \eqref{zetacom}.

\subsection{DFT-vielbein, projectors and the  master derivative \label{SECmaster} }
Here we  review   the semi-covariant differential geometry developed  for double field theory~\cite{Jeon:2011cn}, with the intention  of   incorporating   the higher spin gauge symmetry. Firstly, we recall   the defining property of the DFT-vielbein~(\ref{Vdef}),    
\be
V_{Ap}V^{A}{}_{q}=\eta_{pq}\,.
\label{Vdef2}
\ee
If we view $V_{Ap}$ as an $8\times 4$ matrix and \textit{assume} that   its upper $4\times 4$ block is non-degenerate,  the defining condition~(\ref{Vdef2})  can be generically solved by the following   parametrization,
\be
V_{Ap}=\frac{1}{\sqrt{2}}\left(\ba{c}(e^{-1})_{p}{}^{\mu}\\
(B+e)_{\nu p}\ea\right)\,,
\label{Vpara}
\ee
where, with respect to the aforementioned particular  choice of the section, ${\frac{\partial~~}{\partial\tx_{\mu}}\equiv0}$~\eqref{conventionalSEC},  $e_{\mu}{}^{p}$ corresponds to an ordinary vierbein and thus  sets a Riemannian metric,
\be
g_{\mu\nu}=e_{\mu}{}^{p}e_{\nu}{}^{q}\eta_{pq}\,.
\label{Vpara2}
\ee
Further, in (\ref{Vpara}), we put $B_{\nu p}=B_{\nu\sigma}(e^{-1})_{p}{}^{\sigma}$ with  a skew-symmetric two-form field, $B_{\mu\nu}=-B_{\nu\mu}$. On the other hand, if the upper block is degenerate,  the DFT-vielbein  cannot be parametrized as above. It should be solved in a different manner, and it generically leads to  a `non-Riemannian' stringy   background which does not admit any Riemannian interpretation~\cite{Lee:2013hma,Ko:2015rha}.\footnote{See also \cite{Malek:2013sp} for U-duality manifest generalized metrics.} 
  Unless explicitly stated, hereafter we shall not assume any particular parametrization of the DFT-vielbein like   (\ref{Vpara}).  Only the defining property~(\ref{Vdef2}) needs to be assumed and suffices. \\

The DFT-vielbein produces  a pair of projectors.  Firstly, we set 
\be
P_{AB}:=V_{Ap}V_{B}{}^{p}\,,
\label{defP}
\ee
which is, by construction with  (\ref{Vdef2}),  a symmetric projector,
\be
\ba{ll}
P_{AB}=P_{BA}\,,\quad&\quad
P_{A}{}^{B}P_{B}{}^{C}=P_{A}{}^{C}\,.
\ea
\ee
The complementary  projector is subsequently defined,
\be
\ba{ll}
\brP_{A}{}^{B}:=\delta_{A}{}^{B}-P_{A}{}^{B}\,,\quad&\quad
\brP_{A}{}^{B}\brP_{B}{}^{C}=\brP_{A}{}^{C}\,,
\ea
\ee
such that  both 
$P_{AB}$ and $\brP_{AB}$ are symmetric projectors, being   orthogonal and complete,
\be
\ba{ll}
P_{A}{}^{B}\brP_{B}{}^{C}=0\,,\quad&\quad P_{AB}+\brP_{AB}=\cJ_{AB}\,.
\ea
\ee
Unlike the supersymmetric double field theories~\cite{Jeon:2011sq,Jeon:2012hp},   it is  unnecessary to     introduce   a separate  $\oSpinf$ DFT-vielbein,  $\brV_{A\brp}$,  and to define the orthogonal  projector as its ``square'',  \textit{i.e.~}$\brP_{AB}=\brV_{A}{}^{\brp}\brV_{A\brp}$, which would be  analogous  to (\ref{defP}).   In the current  proposal of the HS-DFT, we make use of only one  spin group, such that $\Psi^{\alpha}$ is  not $\oSpinf$ but $\Spinf$ Majorana spinor.\\

The  two-index projectors further generate `six-index' projectors~\cite{Jeon:2011cn}: with $D=4$,
\be
\ba{ll}
\cP_{ABC}{}^{DEF}:=P_{A}{}^{D}P_{[B}{}^{[E}P_{C]}{}^{F]}+\textstyle{\frac{2}{D-1}}P_{A[B}P_{C]}{}^{[E}P^{F]D}\,,\quad&\quad \cP_{ABC}{}^{DEF}\cP_{DEF}{}^{GHI}=\cP_{ABC}{}^{GHI}\,,
\\
\bar{\cP}_{ABC}{}^{DEF}:=\brP_{A}{}^{D}\brP_{[B}{}^{[E}\brP_{C]}{}^{F]}+\textstyle{\frac{2}{D-1}}\brP_{A[B}\brP_{C]}{}^{[E}\brP^{F]D}\,,\quad&\quad\brcP_{ABC}{}^{DEF}\brcP_{DEF}{}^{GHI}=\brcP_{ABC}{}^{GHI}\,,
\ea
\label{P6}
\ee
which are symmetric and traceless in the following sense,
\be
\ba{lll}
\cP_{ABCDEF}=\cP_{DEFABC}\,,\quad&\quad\cP_{ABCDEF}=\cP_{A[BC]D[EF]}\,,\quad&\quad
P^{AB}\cP_{ABCDEF}=0\,,\\
\brcP_{ABCDEF}=\brcP_{DEFABC}\,,\quad&\quad\brcP_{ABCDEF}=\brcP_{A[BC]D[EF]}\,,\quad&\quad
\brP^{AB}\brcP_{ABCDEF}=0\,.
\ea
\label{symP6}
\ee
~\\

We are now ready to explain the properties of the \textit{master  derivative}~(\ref{Master}), 
\be
\cD_{A}:=\partial_{A}+\Gamma_{A}(x)+\Phi_{A}(x)+\cW_{A}(x,\zeta,\brzeta)\,,
\label{Master2}
\ee
which includes \textit{the  semi-covariant derivative} introduced in \cite{Jeon:2010rw,Jeon:2011cn} for the DFT-diffeomorphisms,
\be
\na_{A}:=\partial_{A}+\Gamma_{A}\,.
\ee
Explicitly, acting on a generic covariant  tensor~(\ref{tcL}), the semi-covariant derivative  reads 
\be
\na_{C}T_{A_{1}A_{2}\cdots A_{n}}
=\partial_{C}T_{A_{1}A_{2}\cdots A_{n}}-\omega_{{\scriptscriptstyle{T\,}}}\Gamma^{B}{}_{BC}T_{A_{1}A_{2}\cdots A_{n}}+
\sum_{i=1}^{n}\,\Gamma_{CA_{i}}{}^{B}T_{A_{1}\cdots A_{i-1}BA_{i+1}\cdots A_{n}}\,,
\label{asemicov}
\ee
and its connection, or the DFT extension of the  Christoffel symbol,  is given by
\be
\ba{ll}
\Gamma_{CAB}=\Gamma_{C[AB]}=&2\left(P\partial_{C}P\brP\right)_{[AB]}
+2\left({{\brP}_{[A}{}^{D}{\brP}_{B]}{}^{E}}-{P_{[A}{}^{D}P_{B]}{}^{E}}\right)\partial_{D}P_{EC}\\
{}&-\textstyle{\frac{4}{D-1}}\left(\brP_{C[A}\brP_{B]}{}^{D}+P_{C[A}P_{B]}{}^{D}\right)\!\left(\partial_{D}d+(P\partial^{E}P\brP)_{[ED]}\right)\,.
\ea
\label{Gammao}
\ee
This expression is  uniquely  fixed  
by requiring  \textit{(i)} the compatibility with the DFT-dilaton and the  projectors,
\be
\ba{lll}
\na_{A}d=-\half e^{2d}\na_{A}(e^{-2d})=\partial_{A}d+\half\Gamma^{B}{}_{BA}=0\,,\quad&\quad
\na_{A}P_{BC}=0\,,\quad&\quad\na_{A}\brP_{BC}=0\,,
\ea
\label{semicovcomp}
\ee
\textit{(ii)} a cyclic  property, 
\be
\Gamma_{ABC}+\Gamma_{BCA}+\Gamma_{CAB}=0\,,
\label{Gsym}
\ee  
and \textit{(iii)} the kernel conditions for the six-index projectors,
\be
\ba{ll}
\cP_{ABC}{}^{DEF}\Gamma_{DEF}=0\,,\quad&\quad\brcP_{ABC}{}^{DEF}\Gamma_{DEF}=0\,.
\ea
\label{kernel}
\ee
The cyclic property~(\ref{Gsym}) corresponds to a torsionless condition, as it ensures that we can freely replace the ordinary derivatives in the definition of the generalized Lie derivative~(\ref{tcL}) by the semi-covariant derivatives,\footnote{In this work we focus on the above torsionless connection~(\ref{Gammao}). Yet, in the `full order' supersymmetric double field theories, in order to ensure the `1.5 formalism', it is necessary to relax (\ref{Gsym}) and  include  torsions which are quadratic in fermions~\cite{Jeon:2011sq,Jeon:2012hp}.} 
\be
\hcL_{X}(\partial)=\hcL_{X}(\na)\,.
\ee
~\\

In general,  the  semi-covariant derivative by itself is  not completely covariant under DFT-diffeomorphisms.\footnote{Nevertheless,  exceptions  exist which include (\ref{semicovcomp}), (\ref{kernel}), (\ref{semicovcomp2}) \textit{etc.} They are completely covariant by themselves, as the  anomalous  terms  in (\ref{diffeoanomal}) vanish automatically   for them.} There is a potential discrepancy between the  actual diffeomorphic transformation and the generalized Lie derivative of the semi-covariant derivative, 
\be
(\delta_{X}{-\hcL_{X}})\na_{C}T_{A_{1}\cdots A_{n}}=
\dis{\sum_{i=1}^{n}2(\cP{+\brcP})_{CA_{i}}{}^{BDEF}
(\partial_{D}\partial_{E}X_{F})\,T_{A_{1}\cdots A_{i-1} BA_{i+1}\cdots A_{n}}\,.}
\label{diffeoanomal}
\ee
However, the diffeomorphic anomaly on the right hand side of the above equality is organized in terms of the six-index projectors, and hence it can be projected out. This explains the notion, `semi-covariance'.  Namely, the characteristic of the semi-covariant derivative is   that, it can be completely covariantized  through  appropriate  contractions with the projectors or the  DFT-vielbein~\cite{Jeon:2010rw,Jeon:2011cn}.  The \textit{completely covariant derivatives}, relevant to the present work,  are from \cite{Jeon:2011cn},\footnote{For the  `updated'  full list of the completely covariant derivatives, we refer  to \cite{Ko:2015rha} (tensorial) and  \cite{Cho:2015lha} (spinorial).} 
\be
\ba{ll}
P_{C}{}^{D}{\brP}_{A_{1}}{}^{B_{1}}\brP_{A_{2}}{}^{B_{2}}\cdots{\brP}_{A_{n}}{}^{B_{n}}
\na_{D}T_{B_{1}B_{2}\cdots B_{n}}\,,\quad&\quad
{\brP}_{C}{}^{D}P_{A_{1}}{}^{B_{1}}P_{A_{2}}{}^{B_{2}}\cdots P_{A_{n}}{}^{B_{n}}
\na_{D}T_{B_{1}B_{2}\cdots B_{n}}\,,\\
P^{AB}{\brP}_{C_{1}}{}^{D_{1}}{\brP}_{C_{2}}{}^{D_{2}}\cdots{\brP}_{C_{n}}{}^{D_{n}}\na_{A}T_{BD_{1}D_{2}\cdots D_{n}}\,,\quad&\quad
\brP^{AB}{P}_{C_{1}}{}^{D_{1}}{P}_{C_{2}}{}^{D_{2}}\cdots{P}_{C_{n}}{}^{D_{n}}\na_{A}T_{BD_{1}D_{2}\cdots D_{n}}\,.
\ea
\label{covT}
\ee

The second connection in the master derivative~(\ref{Master2}) is the  spin connection for the  local Lorentz symmetry of $\Spinf$, 
\be
\Phi_{Apq}=\Phi_{A[pq]}=V^{B}{}_{p}\na_{A}V_{Bq}=V^{B}{}_{p}(\partial_{A}V_{Bq}+\Gamma_{AB}{}^{C}V_{Cq})\,.
\label{Phi}
\ee
It is determined by requiring that the master derivative should be compatible  with the DFT-vielbein (which is  HS gauge singlet),
\be
\cD_{A}V_{Bp}=\partial_{A}V_{Bp}+\Gamma_{AB}{}^{C}V_{Cp}+\Phi_{Ap}{}^{q}V_{Bq}=0\,.
\label{semicovcomp2}
\ee
The master  derivative is also compatible with  the $\Off$ invariant  metric~(\ref{cJ}), the $\Spinf$  Minkowskian flat  metric, the gamma matrices and the charge conjugation matrix:
\be
\ba{llll}
\cD_{A}\cJ_{BC}=\na_{A}\cJ_{BC}=0\,,\quad&\quad\cD_{A}\eta_{pq}=0\,,
\quad&\quad\cD_{A}(\gamma^{p})^{\alpha}{}_{\beta}=0\,,
\quad&\quad\cD_{A}\mathbf{C}_{\alpha\beta}=0\,.
\ea
\ee
In particular, from the compatibility with the gamma matrices, the standard  relation between  the spinorial and the vectorial representations of the spin connection follows
\be
\Phi_{A}{}^{\alpha}{}_{\beta}=\quarter\Phi_{Apq}(\gamma^{pq})^{\alpha}{}_{\beta}\,.
\ee
While the spin connection, $\Phi_{Apq}$, takes good care of the $\Spinf$ local Lorentz symmetry, making  the master derivative always  covariant under the symmetry, it is potentially anomalous for the DFT-diffeomorphisms, like (\ref{diffeoanomal}),
\be
(\delta_{X}-\hcL_{X})\Phi_{Apq}=
2\cP_{Apq}{}^{DEF}\partial_{D}\partial_{E}X_{F}\,.
\label{anomalspinconnection}
\ee
In a similar fashion to (\ref{covT}),  only the following  modules of the spin connection are completely covariant under DFT-diffeomorphisms, 
\be
\ba{lll}
\brP_{A}{}^{B}\Phi_{Bpq}\,,\quad&\quad\Phi_{A[pq}V^{A}{}_{r]}=\Phi_{[pqr]}\,,\quad&\quad
\Phi_{Apq}V^{Ap}=\Phi^{p}{}_{pq}\,.
\ea
\label{covPhi}
\ee 
This implies that, acting on arbitrary  $\Spinf$ spinors like  $\Psi^{\alpha}$ or  $\Psi^{\alpha}_{A}$, of  which the latter   carries   an additional $\Off$ vector index,  the completely covariant `Dirac'  operators  are restricted to be  the followings~\cite{Jeon:2011vx,Jeon:2011sq,Jeon:2012kd,Jeon:2012hp},
\be
\ba{llll}
\gamma^{p}\cD_{p}\Psi=\gamma^{A}\cD_{A}\Psi\,,\quad&\quad
\brP_{A}{}^{B}\cD_{B}\Psi\,,\quad&\quad\brP_{A}{}^{B}\gamma^{p}\cD_{p}\Psi_{B}\,,\quad&\quad\brP^{AB}\cD_{A}\Psi_{B}\,.
\ea
\label{covDirac}
\ee
Explicitly, the  semi-covariant master derivatives of the spinors read 
\be
\ba{l}
{}~~\cD_{A}\Psi(x,\zeta,\brzeta)\,=\partial_{A}\Psi
+\quarter\Phi_{Apq}\gamma^{pq}\Psi
+[\cW_{A},\Psi]_{\star}\,,\\
\cD_{A}\Psi_{B}(x,\zeta,\brzeta)=\partial_{A}\Psi_{B}+\Gamma_{AB}{}^{C}\Psi_{C}
+\quarter\Phi_{Apq}\gamma^{pq}\Psi_{B}
+[\cW_{A},\Psi_{B}]_{\star}\,.
\ea
\label{DPsi}
\ee
Note that, since we postulate in (\ref{Spinftransf}) that the $\Spinf$ local Lorentz symmetry should  act  only on the explicit $\Spinf$ indices of $V_{Ap}$ and $\Psi^{\alpha}$,  without rotating the spinorial  coordinates, \textit{c.f.~}(\ref{modifiedSpinf1}), (\ref{modifiedSpinf2}), the corresponding spin connection, $\Phi_{A}$, also acts  on the explicit $\Spinf$ indices only in (\ref{DPsi}) not on the internal spinorial  coordinates, $\zeta^{\alpha},\brzeta_{\beta}$.\footnote{However, continue to read  (\ref{suggestive}) and discussion  there.}\\

Finally, under the higher spin gauge symmetry~(\ref{HSGS}), 
\be
\ba{lll}
\delta_{\cT}\Psi=[\cT,\Psi]_{\star}\,,\quad&\quad
\delta_{\cT}\Psi_{A}=[\cT,\Psi_{A}]_{\star}\,,\quad&\quad
\delta_{\cT}\cW_{A}=-\cD_{A}\cT\,,
\ea
\label{HSGS2}
\ee
the master derivatives are covariant by themselves,
\be
\ba{ll}
\delta_{\cT}\left(\cD_{A}\Psi\right)=[\cT,\cD_{A}\Psi]_{\star}\,,\quad&\quad
\delta_{\cT}\left(\cD_{A}\Psi_{B}\right)=[\cT,\cD_{A}\Psi_{B}]_{\star}\,.
\ea
\label{HSGS3}
\ee
~\\

\subsection{Curvature and field strength \label{SECcurvature} }
For each gauge potential in the master derivative~(\ref{Master}), we set a  corresponding \textit{semi-covariant curvature} or \textit{semi-covariant field strength}, following \cite{Jeon:2011cn,Jeon:2011kp,Choi:2015bga,Cho:2015lha},
\be
\ba{l}
S_{ABCD}:=\half(R_{ABCD}+R_{CDAB}-\Gamma^{E}{}_{AB}\Gamma_{ECD})\,,\\
\cF^{\Phi}_{ABpq}:=\na_{A}\Phi_{Bpq}-\na_{B}\Phi_{Apq}+\Phi_{Ap}{}^{r}\Phi_{Brq}-\Phi_{Bp}{}^{r}\Phi_{Arq}\,,\\
\cF^{\cW}_{AB}:=\na_{A}\cW_{B}-\na_{B}\cW_{A}+\left[\cW_{A},\cW_{B}\right]_{\star}\,,
\ea
\label{semiC3}
\ee
in which   $R_{ABCD}$ is given by 
\be
R_{CDAB}=\partial_{A}\Gamma_{BCD}-\partial_{B}\Gamma_{ACD}+\Gamma_{AC}{}^{E}\Gamma_{BED}-\Gamma_{BC}{}^{E}\Gamma_{AED}\,.
\label{RABCD}
\ee
By construction, and also due to the section condition, $S_{ABCD}$ satisfies various  identities~\cite{Jeon:2011cn},
\be
\ba{ll}
S_{ABCD}=S_{CDAB}=S_{[AB][CD]}\,,\quad&\quad
S_{ABCD}+S_{BCAD}+S_{CABD}=0\,,\\
(P^{AB}P^{CD}+\brP^{AB}\brP^{CD})S_{ACBD}=0\,,\quad&\quad
P_{A}{}^{C}\brP_{B}{}^{D}(P^{EF}-\brP^{EF})S_{CEDF}=0\,,\\
P_{A}{}^{E}P_{B}{}^{F}\brP_{C}{}^{G}\brP_{D}{}^{H}S_{EFGH}=0\,,\quad&\quad
P_{A}{}^{E}\brP_{B}{}^{F}P_{C}{}^{G}\brP_{D}{}^{H}S_{EFGH}=0\,.
\ea
\label{Sidentities}
\ee
Despite of all these nice properties, $S_{ABCD}$ is not a DFT-diffeomorphism covariant  tensor.  It is widely speculated  that  there is no completely covariant four-index curvature in double field theory which can be constructed out of the geometric  objects only, \textit{i.e.~}the DFT-dilaton and the projectors, \textit{c.f.~}\cite{Jeon:2011cn,Hohm:2011si}.  Yet, like the semi-covariant derivative~(\ref{diffeoanomal}), the  diffeomorphic  anomaly of $S_{ABCD}$ is  governed by the six-index projectors:
\be
(\delta_{X}-\hcL_{X})S_{ABCD}=
2\na_{[A}\Big(\!(\cP{+\brcP})_{B][CD]}{}^{EFG}\partial_{E}\partial_{F}X_{G}\Big)
+2\na_{[C}\Big(\!(\cP{+\brcP})_{D][AB]}{}^{EFG}\partial_{E}\partial_{F}X_{G}\Big)\,.
\label{anomalyS}
\ee
Thus, once again  after being properly contracted  with the projectors, it can produce completely covariant curvatures, such as  two-index `Ricci-type' curvature  and `scalar' curvature~\cite{Jeon:2011cn}:
\be
\ba{ll}
P_{A}{}^{C}\brP_{B}{}^{D}S_{CD}\,,\qquad&\qquad P^{AB}P^{CD}S_{ACBD}=-\brP^{AB}\brP^{CD}S_{ACBD}\,,
\ea
\label{completecovC}
\ee 
for which  we set
\be
S_{AB}=S_{BA}:=S_{ACB}{}^{C}\,.
\ee
~\\

As $\Phi_{Apq}$ is related to $\Gamma_{ABC}$~(\ref{Phi}), so are their curvatures~\cite{Cho:2015lha}. For example, we note
\be
\brP_{A}{}^{B}S_{Bpqr}=\half\brP_{A}{}^{B}\cF^{\Phi}_{Bpqr}\,,
\label{ScF}
\ee
which is however not a  completely covariant tensor.  As for   the  completely covariant curvatures~(\ref{completecovC}), we have 
\be
\ba{ll}
\brP_{A}{}^{B}S_{Bp}=\brP_{A}{}^{B}\cF^{\Phi}_{Bqp}{}^{q}=
\brP_{A}{}^{B}V^{Cq}\cF^{\Phi}_{BCpq}
\,,\quad&\quad S_{pq}{}^{pq}=P^{AB}P^{CD}S_{ACBD}=\cF^{\Phi}_{pq}{}^{pq}+\half \Phi_{Epq}\Phi^{Epq}\,.
\ea
\ee
~\\

In a similar fashion to \eqref{anomalyS},  $\cF^{\cW}_{AB}$ is anomalous  under DFT-diffeomorphisms and also  HS gauge symmetry,
\be
\ba{rll}
\delta_{X}\cF^{\cW}_{AB}&=&\hcL_{X}\!\left(\cF^{\cW}_{AB}\right)-2(\cP+\brP)^{C}{}_{AB}{}^{DEF}\partial_{D}\partial_{[E}X_{F]}\cW_{C}\,,\\
\delta_{\cT}\cF^{\cW}_{AB}&=&\left[\cT,\cF^{\cW}_{AB}\right]_{\star}+\Gamma^{C}{}_{AB}\partial_{C}\cT\,.
\ea
\label{anomalyF}
\ee
The completely covariant field strength  is then given by, \textit{c.f~}\cite{Jeon:2011kp,Choi:2015bga},
\be
(P\cF^{\cW}\brP)_{AB}=
P_{A}{}^{C}\brP_{B}{}^{D}\cF^{\cW}_{CD}=P_{A}{}^{C}\brP_{B}{}^{D}\big(\na_{C}\cW_{D}-\na_{D}\cW_{C}+[\cW_{C},\cW_{D}]_{\star}\big)\,.
\label{comcovF}
\ee
As the two $\Off$ indices of $\cF^{\cW}_{CD}$ are projected  orthogonally,  the above quantity is clearly  covariant under DFT-diffeomorphisms. Further, the section condition implies  
\be
P_{A}{}^{C}\brP_{B}{}^{D}\Gamma^{E}{}_{CD}\partial_{E}=(P\partial^{E}P\brP)_{AB}\partial_{E}=0\,,
\ee
which   immediately implies, with (\ref{anomalyF}),  that  $(P\cF^{\cW}\brP)_{AB}$ is   covariant under   higher spin  gauge symmetry as well~\cite{Jeon:2011kp}: along with (\ref{HSGS2}) and (\ref{HSGS3}),  we have
\be
\delta_{\cT}(P\cF^{\cW}\brP)_{AB}=\left[\cT,(P\cF^{\cW}\brP)_{AB}\right]_{\star}\,.
\ee
~\\

It is  worth while to note that successive applications of the  completely covariant Dirac operators can produce  the  completely covariant   `Ricci-type' curvature as well as the completely covariant field strength.  From the commutator relation,
\be
[\cD_{A},\cD_{B}]\Psi=
\quarter\cF^{\Phi}_{ABpq}\gamma^{pq}\Psi
+[\cF^{\cW}_{AB},\Psi]_{\star}
-\Gamma^{C}{}_{AB}\partial_{C}\Psi\,,
\ee
one can derive, \textit{~c.f.~}\cite{Coimbra:2011nw,Cho:2015lha},
\be
[\gamma^{p}\cD_{p},\brP_{A}{}^{B}\cD_{B}]\Psi=\half \brP_{A}{}^{B} S_{Bp}\gamma^{p}\Psi+\left[(P\cF^{\cW}\brP)_{BA},\gamma^{B}\Psi\right]_{\star}\,,
\label{covcommutator}
\ee
in which each term on the left and the right hand sides of the equality is, from (\ref{covDirac}), completely covariant.  Similarly,  we may also obtain
\be
\ba{l}
(\gamma^{p}\cD_{p})^{2}\Psi+\brP^{AB}\cD_{A}\cD_{B}\Psi+\quarter S_{pq}{}^{pq}\Psi\\
{}=\half[\cF^{\cW}_{pq},\gamma^{pq}\Psi]_{\star}+
2[\cW_{A},\cD^{A}\Psi]_{\star}
+[\cD_{A}\cW^{A},\Psi]_{\star}-\big[\cW_{A},[\cW^{A},\Psi]_{\star}\big]_{\star}\\
{}=\half\left[\cF^{\cW}_{pq}+\Phi_{Apq}\cW^{A},\gamma^{pq}\Psi\right]_{\star}+
\big[\cW^{A},\partial_{A}\Psi+[\cW_{A},\Psi]_{\star}\big]_{\star}+\partial_{A}\left([\cW^{A},\Psi]_{\star}\right)\,.
\ea
\label{GG1}
\ee
Again, each term on the first line is completely covariant, but the other lines are covariant only as a whole expression. Yet,  if $\Psi$ were higher spin gauge singlet,  they would vanish trivially and this would be consistent with the known result~\cite{Coimbra:2011nw,Cho:2015lha}.



\subsection{The action and the BPS equations:  DFT generalization of the Vasiliev   equations\label{SECproposal}}
Our proposed HS-DFT Lagrangian~(\ref{THEACTION}) consists of two parts: $\cL_{\DFT}$  for the `pure' DFT Lagrangian and $\cL_{\sHS}$ for the  `matter' HS Lagrangian.  We recall them here,
\begin{eqnarray}
&&\cL_{\DFT}=e^{-2d}\Big[\,(P^{AB}P^{CD}-\brP^{AB}\brP^{CD})S_{ACBD}-2\Lambda_{\DFT}\,\Big]\,,
\label{pureDFTL2}\\
&&\cL_{\sHS}=\gHS^{-2}\,e^{-2d}\,\Tr\left[P^{AC}\brP^{BD}\cF^{\cW}_{AB}\star\cF^{\cW}_{CD}
+\bar{\Psi}\star\gammaf\gamma^{A}\cD_{A}\Psi-V_{\star}(\Psi)
\right]\,.\label{HSL2}
\end{eqnarray}
In our proposal, we identify  the genuine DFT fields, \textit{i.e.~}$d$ and $V_{Ap}$,  as the geometric objects, while  the $\HSf$-valued fields, \textit{i.e.~}$\cW_{A}$ and $\Psi^{\alpha}$,  are viewed as the matter living on the background the DFT geometry provides.   As stated previously  in section~\ref{SECPROPOSAL},  $\Lambda_{\DFT}$ denotes the DFT version of the cosmological constant~\cite{Jeon:2011cn}, which naturally arises in the Scherk-Schwarz reduction of the  $D{=10}$ half-maximal supersymmetric double field theory with  the `relaxed'  section condition on the twisting matrix~\cite{Geissbuhler:2011mx,Aldazabal:2011nj,Grana:2012rr,Berman:2013cli},\cite{Cho:2015lha}. In particular, we can choose the sign of $\Lambda_{\DFT}$ freely, either positive or negative~\cite{Cho:2015lha}. On account of  the cosmological constant, we let the minimum of the potential  vanish, \textit{i.e.~}$\min[V_{\star}(\Psi)]=0$. The trace in \eqref{HSL2} stands for the  $(\zeta,\brzeta)$-integrals \eqref{trace} over the (real eight-dimensional) internal space. 
This definition of the trace is formal in the sense that we do not discuss subtle issues such as its convergence or the functional class on which it is well defined.\footnote{Such issues become important when solving Vasiliev equations and defining observables in Vasiliev theory but are beyond the scope of this paper.} 
The only property which we actually make use of is that the trace of the star commutator vanishes, neglecting boundary terms, such that
\be
\Tr\left[f\star g\right]=\Tr\left[g\star f\right]\,.
\label{traceproperty}
\ee

In order to derive the full set of the equations of motion, we need to consider the  arbitrary variations of all the elementary   fields, \textit{i.e.}~$\delta d$, $\delta V_{Ap}$, $\delta\cW_{A}$ and $\delta\Psi^{\alpha}$.  Due to the defining property of the DFT-vielbein~(\ref{Vdef}), $\delta V_{Ap}$ is constrained to  meet~\cite{Jeon:2011sq}
\be
\delta V_{Ap}=\brP_{A}{}^{B}\delta V_{Bp}+\delta V_{B[p}V^{B}{}_{q]}V_{A}{}^{q}\,,
\ee
such that
\be
V_{Ap}\delta V^{A}{}_{q}=V_{A[p}\delta V^{A}{}_{q]}\,,
\ee
and  the  variation of $V_{Ap}$ is generated by  an arbitrary $4\times 4$ skew-symmetric matrix, $\Xi_{pq}=-\Xi_{qp}$, and a $\brP$-projected $8\times 4$ matrix, $\Delta_{Ap}$:
\be
\delta V_{Ap}=\brP_{A}{}^{B}\Delta_{Bp}+V_{A}{}^{q}\Xi_{[pq]}\,.
\label{UpsilonXi}
\ee
All together, there are $6+16$ independent degrees of freedom in $\delta V_{Ap}$ which match those of $\delta B_{\mu\nu}$ and $\delta e_{\mu}{}^{p}$ of the Riemannian parametrization~(\ref{Vpara}).

The induced transformation of the pure DFT Lagrangian is rather  well  known, for which it is useful to note that the induced variation of the semi-covariant four-index curvature is  `total derivatives',
\be
\delta S_{ABCD}=\na_{[A}\delta\Gamma_{B]CD}
+\na_{[C}\delta\Gamma_{D]AB}\,.
\label{deltaS}
\ee
Up to total derivatives ($\,\simeq$\,),\, we have,  \textit{c.f. e.g.~}\cite{Jeon:2011sq,Jeon:2012hp},
\be
\delta\cL_{\DFT}\simeq-2\,{\delta d}\,\cL_{\DFT}+4e^{-2d}\delta V^{Ap}\brP_{A}{}^{B}S_{pB}\,.
\label{deltaLDFT}
\ee
We turn to look for the variation of the higher spin  `matter' Lagrangian, $\delta\cL_{\sHS}$. Firstly,  the induced transformation  of the spin connection is~\cite{Jeon:2011sq}
\be
\delta\Phi_{Apq}=\cD_{A}(V^{B}{}_{p}\delta V_{Bq})+V^{B}{}_{p}V^{C}{}_{q}\delta\Gamma_{ABC}\,,
\ee
where and also in (\ref{deltaS}), $\delta\Gamma_{ABC}$ denotes the induced variation of the DFT-Christoffel connection. While it can be  expressed  explicitly  in terms of $\delta V_{Ap}$ and $\delta d$~\cite{Jeon:2011cn},  for the present purpose  of deriving the equations of motion,  the concrete expression    is luckily  unnecessary:   from (\ref{Gsym}) and the triviality of the trace of a star commutator~\eqref{traceproperty}, we  note
\be 
\ba{l}
\delta\Gamma_{ABC}
\Tr\left(\bar{\Psi}\star\gammaf\gamma^{A}\gamma^{BC}\Psi\right)\\
=\delta\Gamma_{[ABC]}
\Tr\left(\bar{\Psi}\star\gammaf\gamma^{ABC}\Psi\right)+
P^{AB}\delta\Gamma_{ABC}(\mathbf{C}\gammaf\gamma^{C})_{\alpha\beta}
\Tr\left([\Psi^{\alpha},\Psi^{\beta}]_{\star}\right)\\
=0\,.
\ea
\ee
We proceed to obtain through  straightforward computations, up to total derivatives,\footnote{\textit{c.f.} appendix of \cite{Jeon:2012hp} for the case of  fermionic dilatinos.}
\be
\ba{ll}
\quarter\cD_{A}(V^{B}{}_{p}\delta V_{Bq})\,\Tr\Big[\bar{\Psi}\star\gammaf\gamma^{A}\gamma^{pq}\Psi\Big]\!\!&
\simeq
-\half\delta V_{Bq}\,\Tr\Big[\bar{\Psi}\star\gammaf\gamma^{ABq}\cD_{A}\Psi\Big]\\
{}&\simeq
\Tr\Big[\delta V_{A}{}^{p}\,\bar{\Psi}\star\gammaf\gamma^{A}\cD_{p}\Psi
-\half \delta V_{Ap}\bar{\Psi}\gamma^{Ap}\star\gammaf
\gamma^{B}\cD_{B}\Psi\Big]\,,
\ea
\ee
and hence
\be
\ba{l}
\delta\Tr\Big[\bar{\Psi}\star\gammaf\gamma^{A}\cD_{A}\Psi\Big]\\
\simeq 
\Tr\Big[\brP^{AB}\delta V_{Ap}\bar{\Psi}\star\gammaf\gamma^{p}\cD_{B}\Psi+
2(\delta\bar{\Psi}-\quarter\delta V_{Ap}\bar{\Psi}\gamma^{Ap})\star\gammaf\gamma^{B}\cD_{B}\Psi-2
\delta\cW_{A}\star\bar{\Psi}\star\gammaf\gamma^{A}\Psi\Big]\,.
\ea
\label{deltaPsikin}
\ee
Further,  from \cite{Park:2015bza} (section 3.3 therein), we have
\be
\ba{l}
\delta
\Tr\left[P^{AC}\brP^{BD}\cF^{\cW}_{AB}\star\cF^{\cW}_{CD}\right]\\
\simeq 4\,\Tr\,\Big[\delta\cW_{A}\star\cD_{B}(P\cF^{\cW}\brP)^{[AB]}\Big]\\
{}\quad~+2V^{Ap}\delta V^{B}{}_{p}\,\Tr\,\Big[
\left\{P\cF^{\cW}(P-\brP)\star\cF^{\cW}\brP\right\}_{AB}+\na_{C}
\left\{(P\cF^{\cW}\brP)_{AB}\star\cW^{C}\right\}\Big]\,.
\ea
\label{deltaPPFF}
\ee
Combining (\ref{deltaPsikin}) and (\ref{deltaPPFF}), we acquire
\be
\ba{ll}
\delta\cL_{\sHS}\simeq&-2{\delta d}\cL_{\sHS}+2\gHS^{-2}e^{-2d}\,\Tr\,\Big[(\delta\bar{\Psi}-\quarter\delta V_{Ap}\bar{\Psi}\gamma^{Ap})\star\gammaf\gamma^{B}\cD_{B}\Psi-\half\delta\bar{\Psi}\star{\mathbf{C}}^{-1}\partial_{\Psi}V_{\star}(\Psi)\Big]\\
{}&+4\gHS^{-2}e^{-2d}\,\Tr\,\Big[
\delta\cW_{A}\star\left\{\cD_{B}(P\cF^{\cW}\brP)^{[AB]}-\half\bar{\Psi}
\star\gammaf\gamma^{A}\Psi\right\}\Big]\\
{}&+2\gHS^{-2}e^{-2d}V^{Ap}\delta V^{B}{}_{p}\,\Tr\,\Big[
\left\{P\cF^{\cW}(P-\brP)\star\cF^{\cW}\brP\right\}_{AB}+\na_{C}
\left\{(P\cF^{\cW}\brP)_{AB}\star\cW^{C}\right\}\\
&\quad~\qquad\qquad\qquad\qquad\qquad
+\half\brP_{B}{}^{C}\bar{\Psi}\star\gammaf\gamma_{A}\cD_{C}\Psi
\Big]\,.
\ea
\label{deltaLHS}
\ee
The full Euler-Lagrange equations  then follow from (\ref{deltaLDFT}) and (\ref{deltaLHS}): for the DFT-dilaton, 
\be
\cL_{\HSDFT}=\cL_{\DFT}+\cL_{\sHS}=0\,;
\label{EOMd2}
\ee
for the DFT-vielbein from the variation,  $\delta V_{Ap}=V_{A}{}^{q}\Xi_{[pq]}$~(\ref{UpsilonXi}),
\be
\bar{\Psi}\star\gamma^{pq}\gammaf\gamma^{A}\cD_{A}\Psi=0\,;
\label{EOMVXi}
\ee
for the DFT-vielbein from the variation, $\delta V_{Ap}=\brP_{A}{}^{B}\Delta_{Bp}$,\footnote{Though not thoroughly   written in terms of $(P\cF^{\cW}\brP)_{AB}$ ~(\ref{comcovF}),   it is straightforward to show from (\ref{anomalyF}) that the expression of (\ref{EOMVDelta}) is completely  covariant under both  DFT-diffeomorphisms and  HS gauge symmetry, \textit{c.f.~}Ref.\cite{Park:2015bza}.}
\be
P_{A}{}^{C}\brP_{B}{}^{D}\left(
S_{CD}+\half\gHS^{-2}\Tr\Big[
\left\{\cF^{\cW}\star(P{-\brP})\cF^{\cW}\right\}_{CD}+
\na_{E}
\left(\cF^{\cW}_{CD}\star\cW^{E}\right)
+\half\bar{\Psi}\star\gammaf\gamma_{C}\cD_{D}\Psi\Big]\right)=0\,;
\label{EOMVDelta}
\ee
for the  HS gauge potential, 
\be
\cD_{B}\left(P\cF^{\cW}\brP\right)^{[AB]}-
\half\bar{\Psi}\star\gammaf\gamma^{A}\Psi=0\,;
\label{EOMWp}
\ee
and lastly for  the bosonic  $\Spinf$ Majorana spinor, 
\be
\gamma^{A}\cD_{A}\Psi-\half\gammaf\mathbf{C}^{-1}\partial_{\Psi}V_{\star}(\Psi)=0\,,
\label{EOMPsi2}
\ee
which actually implies \eqref{EOMVXi} provided the potential, $V_{\star}(\Psi)$, is $\Spinf$ singlet.

From the orthogonality, $\brP_{A}{}^{B}V_{Bp}=0$, the equation of motion of the  HS gauge potential~(\ref{EOMWp}) actually    implies both
\be
\cD_{B}\left(P\cF^{\cW}\brP\right)^{BA}=0\,,
\label{EOMW12}
\ee
and
\be
\cD_{B}\left(P\cF^{\cW}\brP\right)^{AB}-
\bar{\Psi}\star\gammaf\gamma^{A}\star\Psi=0\,.
\label{EOMW22}
\ee
Further, the skew-symmetric property of 
 $(\mathbf{C}\gammaf\gamma^{p})_{\alpha\beta}$ (\ref{mathbfC}) gives an identity,
\be
\bar{\Psi}\star\gammaf\gamma^{p}\Psi=\half \left[\Psi^{\alpha},\Psi^{\beta}\right]_{\star}(\mathbf{C}\gammaf\gamma^{p})_{\alpha\beta}\,,
\ee
and hence we may rewrite (\ref{EOMW22}) as
\be
\cD_{B}\left(P\cF^{\cW}\brP\right)^{AB}-\half
\left[\Psi^{\alpha},\Psi^{\beta}\right]_{\star}(\mathbf{C}\gammaf\gamma^{A})_{\alpha\beta}=0\,.
\ee
~\\

Clearly, all the equations of motion, \eqref{EOMd2}, \eqref{EOMVXi}, \eqref{EOMVDelta}, \eqref{EOMPsi2}, \eqref{EOMW12}, \eqref{EOMW22} are automatically fulfilled, if we assume  the DFT equations of motion, (\ref{HSDFTeqLDFT}), (\ref{HSDFTeq0}), 
\begin{eqnarray}
&&(P^{AB}P^{CD}-\brP^{AB}\brP^{CD})S_{ACBD}-2\Lambda_{\DFT}-\gHS^{-2}\Tr[V_{\star}(\Psi)]=0\,,\label{HSDFTeqLDFT2}\\
&&P_{A}{}^{C}\brP_{B}{}^{D}S_{CD}=0\,,\label{HSDFTeq02}
\end{eqnarray}
and   the first or zeroth order differential    BPS  equations, 
(\ref{HSDFTeq1})
--(\ref{HSDFTeq5}),
\begin{eqnarray}
&&P_{A}{}^{C}\brP_{B}{}^{D}\cF^{\cW}_{CD}(x,\zeta,\brzeta)=0\,,\label{HSDFTeq12}\\
&&\brP_{A}{}^{B}\cD_{B}\Psi(x,\zeta,\brzeta)=0\,,\label{HSDFTeq22}\\
&&\gamma^{A}\cD_{A}\Psi(x,\zeta,\brzeta)=0\,,\label{HSDFTeq32}\\
&& \left[\Psi^{\alpha}(x,\zeta,\brzeta)\,,\,\Psi^{\beta}(x,\zeta,\brzeta)\right]_{\star}(\mathbf{C}\gammaf\gamma^{p})_{\alpha\beta}=0\,,\label{HSDFTeq42}\\
&&\partial_{\Psi}\Tr\left[V_{\star}(\Psi)\right]=0\,.\label{HSDFTeq52}
\end{eqnarray}
~\\
It is worth while to note,  from the commutator relation~(\ref{covcommutator}) that, there is a   mutual consistency among   (\ref{HSDFTeq02}), (\ref{HSDFTeq12}), (\ref{HSDFTeq22}) and    (\ref{HSDFTeq32}). Furthermore, ignoring the adjoint action of the HS gauge potential, the two conditions, (\ref{HSDFTeq22}) and (\ref{HSDFTeq32}), are precisely the supersymmetry transformations of the gravitino and the dilatino respectively  in the half-maximal supersymmetric double field theory~\cite{Jeon:2011sq} (see also \cite{Coimbra:2011nw}). This also supports our nomenclature, `{BPS}'. \\

 
From the $4\times 4$ skew-symmetric completeness relation~(\ref{complete2}), the  algebraic commutator relation~(\ref{HSDFTeq42}) is equivalent to
\be 
[\Psi^{\alpha},\Psi^{\beta}]_{\star}=
-\half(\bar{\Psi}\star\Psi)\mathbf{C}^{-1\alpha\beta}
-\half(\bar{\Psi}\star\gammaf\star\Psi)
(\gammaf \mathbf{C}^{-1})^{\alpha\beta}\,.
\label{HSDFTeq43}
\ee
Thus,  if we let 
\be
\ba{ll}
\cQ_{+}:=
-\bar{\Psi}\star(1+\gammaf)\star\Psi\,,\quad&\quad
\cQ_{-}:=-
\bar{\Psi}\star(1-\gammaf)\star\Psi\,,
\ea
\label{cQdef2}
\ee
satisfying  the reality relation, 
\be
\cQ_{+}=(\cQ_{-})^{\dagger}\,,
\ee
\eqref{HSDFTeq43}  can be  rewritten  as (\ref{PPQQ}), \textit{i.e.} 
\be
\big[\Psi^{\alpha},\Psi^{\beta}\big]_{\star}=\quarter
\big[(1+\gammaf)\mathbf{C}^{-1}\big]^{\alpha\beta}\cQ_{+}
+\quarter\big[(1-\gammaf)\mathbf{C}^{-1}\big]^{\alpha\beta}\cQ_{-}\,.
\ee
~\\

We recall the expansion of the  Majorana spinor  field, $\Psi^{\alpha}$,  in terms of   the internal spinorial  coordinates~(\ref{expansion}),
\be
\dis{\,\Psi^{\alpha}(x,\zeta,\brzeta)\,=\,\sum_{m,n}\frac{1}{m!n!}\,\zeta^{\alpha_{1}}\zeta^{\alpha_{2}}
\cdots\zeta^{\alpha_{m}}\brzeta_{\beta_{1}}\brzeta_{\beta_{2}}
\cdots\brzeta_{\beta_{n}}\Psi^{\alpha}{}_{\alpha_{1}\alpha_{2}\cdots
\alpha_{m}}{}^{\beta_{1}\beta_{2}\cdots\beta_{n}}(x)
\,,}
\ee
and decompose the HS gauge potential as 
\be
\cW_{A}= \textstyle{\frac{1}{4}}\Phi_{Apq}\brzeta\gamma^{pq}\zeta+\cW^{\prime}_{A}\,.
\label{decomcW}
\ee
Then, the master derivative of the spinor field gives, with  \eqref{forPMy},
\be
\ba{ll}
\!\!\cD_{A}\Psi^{\alpha}(x,\zeta,\brzeta)\!\!\!&=\partial_{A}\Psi^{\alpha}(x,\zeta,\brzeta)
+\quarter\Phi_{Apq}(\gamma^{pq})^{\alpha}{}_{\beta}\Psi^{\beta}(x,\zeta,\brzeta)
+\left[\cW_{A},\Psi^{\alpha}(x,\zeta,\brzeta)\right]_{\star}\\
{}&\dis{=
\,\sum_{m,n}~\frac{1}{m!n!}\Big\{\,\zeta^{\alpha_{1}}\zeta^{\alpha_{2}}
\cdots\zeta^{\alpha_{m}}\brzeta_{\beta_{1}}\brzeta_{\beta_{2}}
\cdots\brzeta_{\beta_{n}}\cD_{A}^{\prime}\Psi^{\alpha}{}_{\alpha_{1}\alpha_{2}\cdots
\alpha_{m}}{}^{\beta_{1}\beta_{2}\cdots\beta_{n}}(x)}\\
{}&\qquad\qquad\qquad~\dis{
+\left[\cW^{\prime}_{A}\,,\,\zeta^{\alpha_{1}}\zeta^{\alpha_{2}}
\cdots\zeta^{\alpha_{m}}\brzeta_{\beta_{1}}\brzeta_{\beta_{2}}
\cdots\brzeta_{\beta_{n}}\right]_{\star}
\Psi^{\alpha}{}_{\alpha_{1}\alpha_{2}\cdots
\alpha_{m}}{}^{\beta_{1}\beta_{2}\cdots\beta_{n}}(x)\Big\}}.
\ea
\label{cDPsiprime}
\ee
In the above,  we set for the component field,
\be
\ba{ll}
\cD_{A}^{\prime}\Psi^{\alpha}{}_{\alpha_{1}\cdots
\alpha_{m}}{}^{\beta_{1}\cdots\beta_{n}}\,\equiv&\!\!
\dis{\partial_{A}\Psi^{\alpha}{}_{\alpha_{1}\alpha_{2}\cdots
\alpha_{m}}{}^{\beta_{1}\beta_{2}\cdots\beta_{n}}+
\quarter\Phi_{Apq}(\gamma^{pq})^{\alpha}{}_{\beta}\Psi^{\beta}{}_{\alpha_{1}\cdots
\alpha_{m}}{}^{\beta_{1}\cdots\beta_{n}}}\\
{}&\!\!\!\dis{-\sum_{i=1}^{m}\quarter\Phi_{Apq}(\gamma^{pq})^{\gamma}{}_{\alpha_{i}}\Psi^{\alpha}{}_{\alpha_{1}\cdots\gamma\cdots
\alpha_{m}}{}^{\beta_{1}\cdots\beta_{n}}
+\sum_{j=1}^{n}\quarter\Phi_{Apq}(\gamma^{pq})^{\beta_{j}}{}_{\delta}\Psi^{\alpha}{}_{\alpha_{1}\cdots
\alpha_{m}}{}^{\beta_{1}\cdots\delta\cdots\beta_{n}}\,.}
\ea
\label{suggestive}
\ee
This is a suggestive form, especially with respect to   the possible modification of the $\Spinf$ local Lorentz transformation rule as   (\ref{modifiedSpinf1}) and (\ref{modifiedSpinf2}), since the spin connection now acts equally  on all the spinorial  indices of the component fields.\\  

\subsection{Linear  DFT-dilaton vacuum\label{SEClineard}}
To solve for a vacuum solution, we make an ansatz to put, with (\ref{ReIm}),  (\ref{decomcW}),
\be
\ba{ll}
\mr{\Psi}{}^{\alpha}\equiv m^{\frac{3}{2}}\,\mathbf{Re}(\zeta^{\alpha})=
 \half m^{\frac{3}{2}}(\zeta^{\alpha}+\brzeta_{\beta}{\mathbf{C}}^{-1\beta\alpha})\,, \quad&\quad
\mr{\cW}{}^{\prime}_{A}\equiv0\,,
\ea
\ee 
satisfying, from (\ref{cDPsiprime}),
\be
\ba{lll}
\mr{\cW}_{A}=\textstyle{\frac{1}{4}}\mr{\Phi}_{Apq}\brzeta\gamma^{pq}\zeta\,,\quad&\quad
\mr{\cD}{}_{A}\mr{\Psi}{}^{\alpha}=0\,,\quad&\quad
\big[\mr{\Psi}{}^{\alpha},\mr{\Psi}{}^{\beta}\big]_{\star}=\half m^{3}{\mathbf{C}}^{-1\alpha\beta}\,,
\ea
\label{mrPhi}
\ee
of which the last commutator relation implies
\be
\mr{\cQ}_{+}=\mr{\cQ}_{-}=m^{3}\,.
\ee
Hereafter, the circle `$\,\circ\,$' denotes the vacuum. Since the  HS gauge potential is  a DFT-diffeomorphism covariant vector but the spin connection is  not, \textit{c.f.}~\eqref{anomalspinconnection}, we are now looking for a  vacuum  configuration  which breaks  DFT-diffeomorphisms  spontaneously. As is the case  with the  potentials, (\ref{VYM}) and (\ref{VVasiliev}), we also assume the potential to take the absolute minimum value, $V_{\star}(\mr{\Psi})=\min[V_{\star}]=0$,  when $\Psi=\mr{\Psi}$.

From \eqref{semiC3} and \eqref{ScF}, it is straightforward to obtain
\be
\mr{\cF}{}^{\mr{\cW}}_{AB}=\quarter\mr{\cF}{}^{\mr{\Phi}}_{ABpq}\,\brzeta\gamma^{pq}\zeta\,,
\ee
and
\be
\mr{P}_{A}{}^{C}\mr{\brP}_{B}{}^{D}\mr{\cF}{}^{\mr{\cW}}_{CD}=\half\mr{P}_{A}{}^{C}\mr{\brP}_{B}{}^{D}\mr{S}_{CDpq}\,\brzeta\gamma^{pq}\zeta\,.
\label{PPFPPSM}
\ee
Then, from the identities~\eqref{Sidentities}, including 
\be
\mr{P}_{A}{}^{C}\mr{\brP}_{B}{}^{D}\mr{S}_{CEDF}\mr{P}^{EF}=\mr{P}_{A}{}^{C}\mr{\brP}_{B}{}^{D}\mr{S}_{CEDF}\mr{\brP}{}^{EF}=\half \mr{P}_{A}{}^{C}\mr{\brP}_{B}{}^{D}\mr{S}_{CD}\,,
\ee
we note that \eqref{HSDFTeq12} implies \eqref{HSDFTeq02}. Thus, with (\ref{HSDFTeqLDFT2}) and (\ref{PPFPPSM}),  the remaining  conditions   to fulfill   all the  HS-DFT BPS equations are
\be
\mr{\brP}_{A}{}^{B}\mr{S}_{Bpqr}= 0\,,
\label{lastCON1}
\ee
and
\be
(\mr{P}{}^{AB}\mr{P}{}^{CD}-\mr{\brP}{}^{AB}\mr{\brP}{}^{CD})\mr{S}_{ACBD}-2\Lambda_{\DFT}=0\,.
\label{lastCON2}
\ee
Remarkably,  as we show shortly, backgrounds with  linear DFT-dilaton and constant  DFT-vielbein can satisfy  these two conditions,
\be
\ba{ll}
\mr{d}=\mr{N}_{A}x^{A}\,,
\qquad&\qquad\partial_{A}\mr{V}_{Bp}=0\,.
\ea
\label{lineardflatV}
\ee
Here we have parametrized the linear DFT-dilaton by  an $\ODD$ constant vector, $\mr{N}^{A}$, which should have the mass dimension. Though our main interest lies in the case of $D=4$, as our discussion   holds in  arbitrary spacetime dimensions,  we keep $D$ free for a while. Since $\mr{N}_{A}$ is given by the partial derivative of the DFT-dilaton, 
\be
\mr{N}_{A}=\partial_{A}\mr{d}\,,
\ee
the constant $\ODD$ vector must be null and satisfies the `section condition' for consistency, 
\be
\ba{ll}
 \mr{N}^{A}\mr{N}_{A}=\partial^{A}\mr{d}\,\partial_{A}\mr{d}=0\,,\quad&\quad\mr{N}^{A}\partial_{A}=\partial^{A}\mr{d}\,\partial_{A}=0\,.
\ea
\ee
The corresponding DFT-Christoffel connection and the  spin  connection are all constant,
\be
\ba{ll}
\mr{\Gamma}_{CAB}=-\textstyle{\frac{4}{D-1}}\!\left(\mr{\brP}_{C[A}\mr{\brP}_{B]}{}^{D}+\mr{P}_{C[A}\mr{P}_{B]}{}^{D}\right)\mr{N}_{D}\,,\quad&\quad
\mr{\Phi}_{Apq}=-\textstyle{\frac{4}{D-1}}\mr{V}_{A[p}\,\mr{N}_{q]}\,.
\ea
\ee
In particular, from (\ref{mrPhi}), we have
\be
\ba{llll}
\mr{\cW}_{A}=-\textstyle{\frac{1}{D-1}}\,\brzeta\gamma_{Ap}\zeta\,\mr{N}^{p}\,,\quad&\quad
\mr{\brP}_{A}{}^{B}\mr{\Phi}_{B pq}=0\,,\quad&\quad
\mr{\Phi}_{[pqr]}=0\,,\quad&\quad
\mr{\Phi}{}^{p}{}_{pq}=-2\mr{N}_{q}\,.
\ea
\label{vacuumconnection}
\ee
Finally, from
\be
\mr{R}_{CDAB}=\textstyle{\frac{16}{(D-1)^{2}}}\left[
\mr{P}_{A[D}(\mr{P}\mr{N})_{E]}\,\mr{P}_{B[C}(\mr{P}\mr{N})_{F]}
\,+\,\mr{\brP}_{A[D}(\mr{\brP}\mr{N})_{E]}\,\mr{\brP}_{B[C}(\mr{\brP}\mr{N})_{F]}~-~(A\leftrightarrow B)\right]\cJ^{EF}\,,
\ee
we confirm that the linear DFT-dilaton background indeed solves (\ref{lastCON1}),
\be
\mr{\brP}_{A}{}^{B}\mr{S}_{Bpqr}=0\,,
\ee
giving 
\be
\mr{P}_{A}{}^{C}\mr{\brP}_{B}{}^{D}\mr{S}_{CD}=0\,,
\ee
and produces a constant  scalar curvature,
\be
\mr{S}_{pq}{}^{pq}=\mr{P}{}^{AB}\mr{P}{}^{CD}\mr{S}_{ACBD}=-\mr{\brP}{}^{AB}\mr{\brP}{}^{CD}\mr{S}_{ACBD}
=4\mr{\brP}_{AB}\mr{N}^{A}\mr{N}^{B}=
-4\mr{N}_{p}\mr{N}^{p}\,.
\ee
This fulfills the remaining  last condition~(\ref{lastCON2}), 
provided  the DFT  cosmological constant is matched by
\be
\Lambda_{\DFT}=- 4\mr{N}_{p}\mr{N}^{p}\,.
\ee
~\\

Henceforth, we  consider converting  the DFT frame to the string or Einstein frames, while sacrificing   the manifest $\ODD$ symmetry. Parameterizing   the DFT-vielbein and the DFT-dilaton, either  in terms of   the string framed fields,  $\left\{g_{\mu\nu}, B_{\mu\nu},  \phi\right\}$, through  (\ref{Vpara}), (\ref{Vpara2}) and
\be
e^{-2d}=\sqrt{-g}e^{-2\phi}\,,
\label{dpara}
\ee
or alternatively in terms of the Einstein framed fields,   $\left\{G_{\mu\nu}, B_{\mu\nu},  \Phi\right\}$, by 
\be
\ba{ll}
\phi=\sqrt{\frac{D-2}{8}}\Phi\,,\qquad&\qquad 
g_{\mu\nu}=e^{\sqrt{\frac{2}{D-2}}\Phi}G_{\mu\nu}\,,
\ea
\label{DIC}
\ee
the pure DFT Lagrangian~(\ref{pureDFTL2}) gives, with $H={\rm d}B$ and  up to total derivatives ($\simeq$), 
\be
\sqrt{-g}e^{-2\phi}\!\left(R_{g}+4(\partial\phi)_{g}^{2}-\textstyle{\frac{1}{12}}H^{2}_{g}-2\Lambda_{\DFT}\right)\simeq
\sqrt{-G}\!\left(\!R_{G}-\half(\partial\Phi)_{G}^{2}-\textstyle{\frac{1}{12}}e^{\sqrt{\frac{8}{D-2}}\Phi}H^{2}_{G}-2\Lambda_{\DFT} e^{\sqrt{\frac{2}{D-2}}\Phi}\!\right),
\label{sEL}
\ee
in which the superscripts, $g$ and $G$, indicate which metric is  used. \\

Now, especially for the linear DFT-dilaton vacuum~(\ref{lineardflatV}), if we assume $\mr{N}^{p}$ is a space-like $D$-dimensional  vector and hence the DFT cosmological constant is negative,
\be
\Lambda_{\DFT}=-4\mr{N}_{p}\mr{N}^{p}<0\,,
\ee 
we may write the solution as 
\be
\ba{ll}
\phi=\sqrt{-\Lambda_{\DFT}/2}\,x^{D-1}\,,\quad&\quad
g_{\mu\nu}\rd x^{\mu}\rd x^{\nu}=\eta_{\mu\nu}\rd x^{\mu}\rd x^{\nu}\,,
\ea
\ee
where $x^{D-1}$ denotes the last spatial coordinate, $x^{\mu}=(x^{0},x^{1},\cdots, x^{D-1})$. \\

With the dictionary~(\ref{DIC}),   the above vacuum solution  then  corresponds to  the known $(D{-2})$-brane background  obtained in  \cite{Lu:1996rhb}   in the  Einstein  frame:  introducing one new coordinate,  $x^{D-1}\rightarrow z$, satisfying
\be
\ba{ll}
\sqrt{-2\Lambda_{\DFT}}\, z = e^{\sqrt{-2\Lambda_{\DFT}}\,x^{D-1}}\equiv K\,,\quad&\quad 
\rd z=K\,\rd x^{D-1}\,,
\ea
\ee
the linear DFT-dilaton  vacuum  in the DFT/string frame becomes  the brane configuration    in the Einstein frame, 
\be
\ba{ll}
e^{\Phi}=K^{\sqrt{\frac{2}{D-2}}}\,,\quad&\quad
G_{\mu\nu}\rd x^{\mu}\rd x^{\nu}=K^{-\frac{2}{D-2}}\eta^{\prime}_{\mu^{\prime}\nu^{\prime}}\rd x^{\mu^{\prime}}\rd x^{\nu^{\prime}}+K^{-2\left(\frac{D-1}{D-2}\right)}\rd z^{2}\,.
\ea
\ee
~\\

\subsection{Truncation to the bosonic   Vasiliev HS equations}\label{trunc}
In order to truncate our proposal of the  HS-DFT  BPS   equations, (\ref{HSDFTeqLDFT})\,--\,(\ref{HSDFTeq5}), and to derive the bosonic  four-dimensional  Vasiliev HS equations, it is necessary to  impose some  extra conditions. \\

Firstly,   we constrain   the higher spin gauge potential  to meet, \textit{c.f.~}\cite{Choi:2015bga},
\be
\ba{ll}
\cW^{A}\partial_{A}\equiv0\,,\qquad&\qquad
\cW^{A}\cW_{A}\equiv0\,.
\ea
\label{cWconstraint}
\ee
These imply with the section condition~(\ref{seccon}),
\be
(\partial_{A}+\cW_{A})(\partial^{A}+\cW^{A})\equiv0\,.
\ee
For consistency, under all the symmetry transformations, including the $\Off$ T-duality rotations,  DFT-diffeomorphisms,  and the  HS gauge symmetry, the  above constraints~(\ref{cWconstraint}) are well preserved, such as  $(\delta_{X}\cW^{A})\partial_{A}\equiv0$~(\ref{GLieDeriv}), $\,(\delta_{\cT}\cW^{A})\partial_{A}\equiv0$~(\ref{HSGS}), \textit{etc.}  \\

We fix the section as (\ref{conventionalSEC}), and consequently solve  the   constraints~(\ref{cWconstraint}),
\be
\ba{ll}
\partial_{A}=\left(\frac{\partial~~}{\partial \tx_{\mu}}\,,\,\frac{\partial~~}{\partial x^{\mu}}\right)\equiv\left(0\,,\,\frac{\partial~~}{\partial x^{\mu}}\right)\,,\quad&\quad
\cW_{A}=\left(\tilde{W}^{\mu}\,,\,W_{\mu}\right)\equiv\left(0\,,\,W_{\mu}\right)\,.
\ea
\label{parcW}
\ee
We proceed to  parametrize the DFT-dilaton and the DFT-vilebein in terms of the string framed   Riemannian fields, $\{\phi, B_{\mu\nu}, e_{\mu}{}^{p}, g_{\mu\nu}\}$, as
 (\ref{Vpara}), (\ref{Vpara2}) and (\ref{dpara}).  Then the  DFT equations of motion, (\ref{HSDFTeqLDFT}), (\ref{HSDFTeq0}), correspond to nothing but  the Euler-Lagrange equations of the   string framed Lagrangian of (\ref{sEL}), up to the potential term:  
\be
\ba{ll}
R_{\mu\nu}+2\trd_{\mu}\partial_{\nu}\phi-\quarter H_{\mu\rho\sigma}H_{\nu}{}^{\rho\sigma}=0\,,\quad&\quad
\trd^{\lambda}H_{\lambda\mu\nu}
-2(\partial^{\lambda}\phi)H_{\lambda\mu\nu}=0\,,\\
\multicolumn{2}{c}{
e^{+2d}\cL_{\DFT}-
\gHS^{-2}\Tr\left[V_{\star}(\Psi)\right]=
R+4\Box\phi-4\partial_{\mu}\phi\partial^{\mu}\phi-\textstyle{\frac{1}{12}}H_{\mu\nu\rho}H^{\mu\nu\rho}-2\Lambda_{\DFT}-
\gHS^{-2}\Tr\left[V_{\star}(\Psi)\right]=0\,.}
\ea
\ee 
On the other hand,  the remaining  first or zeroth order differential BPS conditions, (\ref{HSDFTeq1})
--(\ref{HSDFTeq4}), become
\begin{eqnarray}
&&\partial_{\mu}W_{\nu}-\partial_{\nu}W_{\mu}+\left[W_{\mu},W_{\nu}\right]_{\star}=0\,,\label{dftV1}\\
&&\partial_{\mu}\Psi+\quarter\omega_{\mu pq}\gamma^{pq}\Psi+\textstyle{\frac{1}{8}}H_{\mu pq}\gamma^{pq}\Psi+\left[W_{\mu},\Psi\right]_{\star}=0\,,\label{dftV2}\\
&&{}\gamma^{\mu}\left(\partial_{\mu}\Psi+\quarter\omega_{\mu pq}\gamma^{pq}\Psi+\textstyle{\frac{1}{24}}H_{\mu pq}\gamma^{pq}\Psi-\partial_{\mu}\phi\,\Psi+\left[W_{\mu},\Psi\right]_{\star}\right)=0\,,\label{dftV3}\\
&&{}\big[\Psi^{\alpha},\Psi^{\beta}\big]_{\star}=\quarter
\big\{(1+\gammaf)\mathbf{C}^{-1}\big\}^{\alpha\beta}\cQ_{+}
+\quarter\big\{(1-\gammaf)\mathbf{C}^{-1}\big\}^{\alpha\beta}
\cQ_{-}\,,\label{dftV4}\\
&&{}\partial_{\Psi}\Tr\left[V_{\star}(\Psi)\right]=0\,,\label{dftV5}
\end{eqnarray}
where $\omega_{\mu pq}=(e^{-1})_{p}{}^{\nu}\trd_{\mu}e_{\nu q}$ denotes the standard  spin connection in supergravity, \textit{e.g.}~\cite{Coimbra:2011nw}, and $\cQ_{\pm}$ are defined in (\ref{cQdef}).  Combining (\ref{dftV2}) and (\ref{dftV3}), we have an algebraic relation, 
\be
\left(\gamma^{\mu}\partial_{\mu}\phi+\textstyle{\frac{1}{12}}\gamma^{\mu\nu\rho}H_{\mu\nu\rho}\right)\Psi=0\,.
\ee

Secondly, we let the DFT-dilaton  and  the DFT-vielbein  all trivial, or constant. This breaks the  local $\Spinf$ Lorentz symmetry  to a global symmetry, and sets  the DFT-Christoffel connection, the local Lorentz spin connection, and the semi-covariant four-index curvature all trivial,
\be
\ba{lll}
\Gamma_{ABC}\equiv0\,,\quad&\quad\Phi_{Apq}\equiv0\,,\quad&\quad
S_{ABCD}\equiv0\,,
\ea
\ee
such that, for the Riemannian fields we have
\be
\ba{llll}
\omega_{\mu pq}\equiv0\,,\quad&\quad R_{\mu\nu}\equiv0\,, \quad&\quad H_{\lambda\mu\nu}\equiv0\,,
\quad&\quad\partial_{\mu}\phi\equiv0\,.
\ea
\ee
Basically, it eliminates any trace of DFT, solving the pure DFT equations of motion, (\ref{HSDFTeqLDFT}) and (\ref{HSDFTeq0}),  in a trivial manner,  with  the   vanishing  cosmological constant, $\Lambda_{\DFT}\equiv0$.\\

Thirdly,  we assume the bosonic truncation~(\ref{bTrunc}), such that  the HS gauge potential and the bosonic Majorana spinor should meet
\be
\ba{ll}
W_{\mu}(x,-\zeta,-\brzeta)=+W_{\mu}(x,\zeta,\brzeta)\,,\quad&\quad
\Psi^{\alpha}(x,-\zeta,-\brzeta)=-\Psi^{\alpha}(x,\zeta,\brzeta)\,.
\ea
\label{bTrunc2}
\ee

We proceed to convert the four-component $\Spinf$ spinors to two-component Weyl spinors. For this, we put the gamma five matrix into a diagonal form,
\be
\gammaf=\left(\ba{cc}\delta^{\dalpha}{}_{\dbeta}&0\\0&-\delta_{\udalpha}{}^{\udbeta}\ea\right)\,,
\ee
and   decompose the  Majorana and pseudo-Majorana  spinors into  chiral and anti-chiral Weyl spinors,   
\be
\ba{lll}
\Psi^{\alpha}=i\half m^{\frac{3}{2}}\left(\ba{c}s^{\dalpha}\\ \brs_{\udalpha}\ea\right)\,,\quad&\quad
\mathbf{Re}(\zeta)=\zeta_{+}^{\alpha}=\half\left(\ba{c}y^{\dalpha}\\ \bry_{\udalpha}\ea\right)\,,\quad&\quad
\mathbf{Im}(\zeta)=\zeta_{-}^{\alpha}=\half\left(\ba{c}z^{\dalpha}\\ \brz_{\udalpha}\ea\right)\,.
\ea
\ee
Hereafter, the top-dotted and the bottom-dotted indices, $\dalpha,\dbeta=1,2$ and $\udalpha,\udbeta=1,2$, are the chiral and  anti-chiral Weyl spinorial indices. We also employ the following explicit  representation of the gamma matrices,
\be
\gamma^{p}=\left(\ba{cc}0&(\sigma^{p})^{\dalpha\udbeta}\\
(\bar{\sigma}^{p})_{\udalpha\dbeta}&0\ea\right)\,,
\ee
as well as  the $\mathbf{A},\mathbf{B},\mathbf{C}$ matrices, (\ref{mathbfA}), (\ref{mathbfC}), (\ref{BM}),
\be
\ba{lll}
\mathbf{A}=\left(\ba{cc}0&-i\delta_{\udalpha}{}^{\udbeta}\\
+i\delta^{\dalpha}{}_{\dbeta}&0\ea\right)\,,\quad&\quad
\mathbf{B}=\left(\ba{cc}0&-\bar{\epsilon}^{-1\,\udalpha\udbeta}\\
-\epsilon_{\dalpha\dbeta}&0\ea\right)\,,\quad&\quad
\mathbf{C}=\left(\ba{cc}+i\epsilon_{\dalpha\dbeta}&0~\\
0&-i\bar{\epsilon}^{-1\,\udalpha\udbeta}\ea\right)\,,
\ea
\label{ggAABBCC}
\ee
where, as $2\times 2$ matrices,  $\,\sigma^{0}$ and  $-\bar{\sigma}^{0}\,$  are  identity matrices,
\be
\sigma^{0}=-\bar{\sigma}^{0}=\left(\ba{cc}1&0\\0&1\ea\right)\,,
\ee
$\sigma^{i}{=\bar{\sigma}^{i}}$ ($i=1,2,3$) are the  Pauli matrices,
\be
\ba{lll}
\sigma^{1}=\bar{\sigma}^{1}=
\left(\ba{cc}0&1\\1&0\ea\right)\,,\quad&\quad
\sigma^{2}=\bar{\sigma}^{2}=\left(\ba{cc}0&-i\\i&0\ea\right)\,,\quad&\quad
\sigma^{3}=\bar{\sigma}^{3}=\left(\ba{cc}1&0\\0&-1\ea\right)\,,
\ea
\ee
and $\,\epsilon_{\dalpha\dbeta}$, $\,\bar{\epsilon}_{\udalpha\udbeta}$ correspond to the  usual skew-symmetric $2\times2$ matrices,
\be
\epsilon=\bar{\epsilon}=-\epsilon^{-1}=-\bar{\epsilon}^{-1}
=\left(\ba{cc}0&1\\-1&0\ea\right)\,,
\ee
satisfying
\be
(\bar{\sigma}^{p})_{\udalpha\dbeta}=\epsilon_{\dalpha\dgamma}
(\sigma^{p})^{\dgamma\uddelta}\bar{\epsilon}_{\uddelta\dbeta}\,.
\ee
It follows that the Majorana and the pseudo-Majorana  conditions, (\ref{Majorana}), (\ref{MpseudoM}),  agree with the Vasiliev's reality conditions,
\be
\ba{lll}
s_{\dalpha}
=s^{\dbeta}\epsilon_{\dbeta\dalpha}
=-(\brs_{\udalpha})^{\dagger}\,,\quad&\quad
y_{\dalpha}=y^{\dbeta}\epsilon_{\dbeta\dalpha}
=(\bry_{\udalpha})^{\dagger}\,,\quad&\quad
z_{\dalpha}=z^{\dbeta}\epsilon_{\dbeta\dalpha}
=-(\brz_{\udalpha})^{\dagger}\,.
\ea
\ee
Basically, under the complex conjugation, the top-dotted index and the bottom-dotted index  flip to each other, while their positions can be  raised or lowered   by  $\epsilon$, $\bar{\epsilon}$ and the inverses only.\footnote{This differs from the four-component spinorial  convention~(\ref{Majorana}).} Further,   the   star commutator relations read in terms of the Weyl spinor variables, from (\ref{zetacom2}), 
\be
\ba{llll}
[y_{\dalpha},y_{\dbeta}]_{\star}=+2i\epsilon_{\dalpha\dbeta}\,,
\quad&\quad
[\bry_{\udalpha},\bry_{\udbeta}]_{\star}=+2i\bar{\epsilon}_{\udalpha\udbeta}\,,\quad&\quad
{}[z_{\dalpha},z_{\dbeta}]_{\star}=-2i\epsilon_{\dalpha\dbeta}\,,
\quad&\quad
[\brz_{\udalpha},\brz_{\udbeta}]_{\star}=-2i\bar{\epsilon}_{\udalpha\udbeta}\,.
\ea
\ee

The remaining nontrivial  HS-DFT BPS equations, (\ref{dftV1}), (\ref{dftV2}), (\ref{dftV3}), (\ref{dftV4}),  then  reduce to 
\be
\ba{c}
\partial_{\mu}W_{\nu}-\partial_{\nu}W_{\mu}+\left[W_{\mu},W_{\nu}\right]_{\star}=0\,,\\
\partial_{\mu}s_{\dalpha}+\left[W_{\mu},
s_{\dalpha}\right]_{\star}=0\,,\quad\qquad
\partial_{\mu}\brs_{\udalpha}+\left[W_{\mu},
\brs_{\udalpha}\right]_{\star}=0\,,\\
{}[s_{\dalpha},s_{\dbeta}]_{\star}
=-2i\epsilon_{\dalpha\dbeta}\,m^{-3}\cQ_{+}\,,\quad\qquad
{}[s_{\dalpha},\brs_{\udbeta}]_{\star}=0\,,\quad\qquad
[\brs_{\udalpha},\brs_{\udbeta}]_{\star}=
-2i\brepsilon_{\udalpha\udbeta}\,m^{-3}\cQ_{-}\,,
\ea
\ee
where $\cQ_{+}$ and $\cQ_{-}$ are the quantities defined in (\ref{cQdef}) which now read in terms of the Weyl spinors, 
\be
\ba{ll}
\cQ_{+}=i\half m^{3} s_{\dalpha}\star s^{\dalpha}\,,\quad&\quad
\cQ_{-}=i\half m^{3} \brs_{\udalpha}\star\brs^{\udalpha}=\cQ_{+}^{\dagger}\,.
\ea
\ee
The  extra conditions   to fully achieve the bosonic  Vasiliev HS equations are then, 
\be
\ba{ll}
\left\{s_{\dalpha}\,,\,\cQ_{+}\right\}_{\star}=2m^{3}s_{\dalpha}\,,\qquad&\qquad
\left\{\brs_{\udalpha}\,,\,\cQ_{-}\right\}_{\star}
=2m^{3}\brs_{\udalpha}\,,
\ea
\label{scQ}
\ee
and, with the inner Klein operators~\eqref{innerKlein},
\be
\cQ_{+}-m^{3}=\left(\cQ_{-}-m^{3}\right)\star\bfk\star\brbfk\,.
\label{Qreality2}
\ee
The algebraic  conditions of (\ref{scQ}) are equivalent to the four-component expressions  of (\ref{PsicQ}), as well as to\footnote{Alternative to the deformed oscillator relation~(\ref{dor}), the commutator relations~(\ref{YMBPS}) become in terms of the Weyl spinors,
\[\ba{ll}
\big[\,s_{\dalpha}\,,\, s_{\dbeta}s^{\dbeta}\,\big]=0\,,\quad&\quad
\big[\,\brs_{\udalpha}\,,\, \brs_{\udbeta}\brs^{\udbeta}\,\big]=0\,.
\ea
\]}
\be
\ba{ll}
i\quarter s_{\dalpha}\star s_{\dbeta}\star s^{\dalpha}= s_{\dbeta}\,,\qquad&\qquad
i\quarter \brs_{\udalpha}\star\brs_{\udbeta}\star\brs^{\udalpha}= \brs_{\udbeta}\,.
\ea
\label{dor}
\ee
Further, they provide (though not the most general) solutions to  the very last BPS equation~(\ref{dftV5}), once we choose the ``deformed oscillator" potential, $V^{\Vasiliev}_{\star}(\Psi)$ in (\ref{VVasiliev}).  Especially, the cases of   $\cQ_{+}=\cQ_{-}$ (real)   or  $\cQ_{+}{-m^{3}}=-(\cQ_{-}{-m^{3}})$ (pure imaginary) correspond to A or B model  respectively, in the sense of  Ref.\cite{Sezgin:2003pt}. To summarize, we have obtained the bosonic Vasiliev equations~\cite{Vasiliev:1992av}, written in the concise form\footnote{This equivalent form of the equations allows us to omit the explicit writing of the inner Klein operators~(\ref{innerKlein}) --\,although they turn out to be useful in order to make explicit contact with Fronsdal equations in the perturbative expansion\,-- and the Weyl zero-form (which can be reconstructed from $\Psi$).} presented in \cite{Alkalaev:2014nsa}.\\
\\

Finally, it is worth while to note that, upon the sectioning condition for the HS gauge potential~(\ref{parcW}), if we set $\phi$ and $B$-field trivial, our proposed HS-DFT functional~(\ref{THEACTION}) reduces to a `undoubled' (Riemannian) gravity action,
\be
\dis{\int\rd^{4}x~e\!\left[R+\gHS^{-2}\,\Tr\left\{-\quarter F_{\mu\nu}\star F^{\mu\nu}+\textstyle{\frac{1}{\sqrt{2}}}\bar{\Psi}\star\gammaf\gamma^{\mu}D_{\mu}\Psi-V_{\star}(\Psi)\right\}
\right]\,,}
\label{Raction}
\ee
where we set $F_{\mu\nu}=\partial_{\mu}W_{\nu}-\partial_{\nu}W_{\mu}+\left[W_{\mu},W_{\nu}\right]_{\star}$ \,and\, $D_{\mu}=\trd_{\mu}+\omega_{\mu}+W_{\mu}$ denotes the Riemannian master derivative, such that   $D_{\mu}\Psi=\partial_{\mu}\Psi+\quarter\omega_{\mu pq}\gamma^{pq}\Psi+\left[W_{\mu},\Psi\right]_{\star}\,$,   $\,D_{\mu}e_{\nu}{}^{p}=\trd_{\mu}e_{\nu}{}^{p}+\omega_{\mu}{}^{p}{}_{q}e_{\nu}{}^{q}=0\,$.  \\

The remarkable fact is then  that the following four `BPS' conditions,
\be
\ba{llll}
F_{\mu\nu}=0\,,\quad&\quad
D_{\mu}\Psi=0\,,\quad&\quad
 \big[\Psi^{\alpha},\Psi^{\beta}\big]_{\star}(\mathbf{C}\gammaf\gamma^{\nu})_{\alpha\beta}=0\,,\quad&\quad
\partial_{\Psi}\Tr\left[V_{\star}(\Psi)\right]=0\,,
\ea
\ee 
supplemented by an Einstein manifold relation,
\be
R_{\mu\nu}=\half g_{\mu\nu}\,\gHS^{-2}\,\Tr\left[V_{\star}(\Psi)\right]\,,
\ee   
can automatically solve the full set of the Euler-Lagrange equations of the above action which are explicitly,
\be
\ba{l}
R_{\mu\nu}-\half g_{\mu\nu}R\\
=\half\gHS^{-2}\,\Tr\!\left[F_{\mu\lambda}\star F_{\nu}{}^{\lambda}-\quarter g_{\mu\nu}F_{\kappa\lambda}\star F^{\kappa\lambda}
-\textstyle{\frac{1}{\sqrt{2}}}\bar{\Psi}\star\gammaf\left\{\gamma_{(\mu}D_{\nu)}-g_{\mu\nu}\gamma^{\lambda}D_{\lambda}\right\}\Psi-g_{\mu\nu}V_{\star}(\Psi)
\right]\,,\\
D_{\mu}F^{\mu\nu}
=\textstyle{\frac{1}{\sqrt{2}}}\big[\Psi^{\alpha},\Psi^{\beta}\big]_{\star}(\mathbf{C}\gammaf\gamma^{\nu})_{\alpha\beta}
\,,\\
\gamma^{\mu}D_{\mu}\Psi-\textstyle{\frac{1}{\sqrt{2}}}\gammaf\mathbf{C}^{-1}\partial_{\Psi}\Tr\left[V_{\star}(\Psi)\right]=0\,.
\ea
\label{RVeq}
\ee
The present paper has dealt  with its DFT generalization.\\
~\\
\newpage

\section{Comments\label{SECCOMMENTS}}

In this work, we have constructed   a higher spin double field theory  which is an extension of DFT by the $\HSf$-valued fields present in Vasiliev equations.  We  have proposed an invariant functional and derived the corresponding Euler-Lagrange equations, in terms of  the semi-covariant geometry which manifests all the symmetries.  Further,  we have identified a minimal set of BPS conditions which automatically solve all the equations of motion. The conditions  reduce to the four-dimensional bosonic Vasiliev HS equations,  once  extra algebraic conditions, (\ref{Wseccon}), (\ref{PsicQ}) and (\ref{Qreality})   are imposed.  By introducing the so-called outer Klein operators and  relaxing some parity constraint, it should be possible to extend the analysis to the supersymmetric Vasiliev  equations. 
Besides, without the extra constraints, while employing the ``Yang-Mills" potential~(\ref{VYM}), our proposal might provide some bridge between open string and higher spin theory.
\\

The linear dilaton  vacuum solution~\eqref{lineardvacuum}  derived in section~\ref{SEClineard} is valid for arbitrary values  of the DFT cosmological constant, including the trivial case of $\Lambda_{\DFT}=0$. As can be seen easily  from \eqref{vacuumconnection}, the vacuum  does not satisfy the  extra condition of \eqref{Wseccon}. That is  to say, the linear dilaton vacuum  is a genuine DFT configuration  which cannot be realized in the undoubled Vasiliev HS theory. Surely, it differs from the known \textit{AdS} solution of the Vasiliev equations which utilizes the $\so(2,3)$ algebra of (\ref{so23}).  \\

While the ``deformed oscillator''  
potential~(\ref{VVasiliev}) seems to be a proper choice of the potential, as the deformed oscillator relations~(\ref{dor}) solve  the algebraic BPS condition~\eqref{HSDFTeq5}, the alternative ``Yang-Mills'' potential~(\ref{VYM}) also appears to deserve further investigations: in the low  energy  limit,     we have  $\cQ_{\pm}\rightarrow m^{3}$  and hence the deformed oscillator relations can be dynamically achieved. As the higher spin gauge theory is expected to arise in a tensionless limit of string theory,  the incorporation of a Brout-Englert-Higgs mechanism may give mass to the higher spin fields and  make contact with  string field theory. \\

The spin groups in (supersymmetric) double field theory are known to be twofold which  reflects the existing two separate locally inertial frames for the left and the right closed string modes~\cite{Duff:1986ne}. Yet, in the present work we have focused on one of the two spin groups  and extended it to include the higher spin gauge symmetry, $\HSf$. It will be of interest  to extend the twofold spin groups to realize the `doubled' higher spin algebras~\textit{c.f.~}\cite{Boulanger:2015kfa},
\be
\ba{lll}
\Spin(1,D{-1})\times\Spin(D{-1},1)~~&\longrightarrow&~~
\Spin(1,D{-1})\times\HS(D)\times\Spin(D{-1},1)\times\overline{\HS}(D)\,.
\ea
\label{extension2}
\ee

\indent Instead of  the Riemannian parametrization of the DFT-vielbein~(\ref{Vpara}),  we may  consider an  ansatz where the upper $4{\times 4}$ block of $V_{Ap}$ is degenerate, and hence does not admit any Riemannian geometry interpretation~\cite{Lee:2013hma}.   Such a non-Riemannian geometry was shown in \cite{Ko:2015rha} to  provide a  genuine  stringy  background   for the non-relativistic closed string theory \textit{a la} Gomis and Ooguri~\cite{Gomis:2000bd}. In this way, we might be able to obtain a non-relativistic higher spin gravity.\\

On the one hand, inspired by the conjectured relation between Vasiliev theory and the tensionless limit of \textit{open} strings, we have constructed a HS-DFT in which the $\HSf$-valued fields present in Vasiliev equations are treated as `matter' minimally coupled to DFT, while the DFT is treated as gravity and all the NS-NS massless fields are $\HSf$-singlets. On the other hand, in the light of the relation between higher spin gravity and the tensionless limit of \textit{closed} strings, it would be highly desirable  to build a HS-DFT in a much more ambitious sense: a fully unified theory with a single massless spin-two field transforming both under DFT and HS gauge transformations, \textit{i.e.~}a HS-DFT \textit{gravity}. Certainly this goes beyond the scope of the present work.\\ ~\\~\\





\section*{Acknowledgements} 
XB acknowledges  Asia Pacific Center for Theoretical Physics (APCTP) in Pohang, Korea, 
for hospitality where part of this work was done during the program, ``Duality and Novel Geometry in M-theory''.  JHP is grateful to the members of the Laboratoire de Math\'ematiques et Physique Th\'eorique (LMPT) at the Universit\'{e} Francois  Rabelais de Tours, France,   the Mainz Institute for Theoretical Physics (MITP), Germany, and Tohoku University, Japan,  for their warm hospitalities   while  writing this manuscript.\\

We wish to thank the anonymous referee of JHEP for constructive suggestions, especially pointing out the absence of (\ref{Qreality}) in the first submission. We also thank S. Didenko for pointing this out independently while the first version of the paper had been already submitted to JHEP.\\

The research of XB was supported by the Russian Science Foundation grant 14-42-00047 in association with the Lebedev Physical Institute.
The work of JHP was  supported by  the National Research Foundation of Korea   through  the Grants   2013R1A1A1A05005747, 2015K1A3A1A21000302 and 2016R1D1A1B0101519. \\  
\newpage


\appendix

\begin{center}
\Large{\bf{{APPENDIX}}}
\end{center}
~\\
~\\
\section{Star product and Wick Theorem\label{sectionAppendix}}
Here, explicitly   we present   our own  proofs for the various properties of the star product.\\

\subsection{Equivalence of \eqref{star1} and \eqref{star2}\label{Appendixequivalence}}
The equivalence of the two expressions for the star product, (\ref{star1}) and (\ref{star2}) can be directly established:
\be
\ba{l}
\textstyle{\frac{1}{(2\pi)^{4}}}\dis{\int\rd^{4} \lambda_{+}\int\rd^{4}\rho_{+}~~e^{\bar{\lambda}_{+}\rho_{+}}f(x,\zeta+\lambda_{+},\brzeta)g(x,\zeta,\brzeta+\bar{\rho}_{+})}\\
=\textstyle{\frac{1}{(2\pi)^{4}}}\dis{\int\rd^{4} \lambda_{+}\int\rd^{4}\rho_{+}~~e^{-\brrho_{+}\lambda_{+}}f(x,\zeta+\lambda_{+},\brzeta)\exp\left(\brrho_{+\alpha}\frac{\partial~}{\partial \brzeta_{\alpha}}\right)g(x,\zeta,\brzeta)}\\
{}=\textstyle{\frac{1}{(2\pi)^{4}}}\dis{\int\rd^{4} \lambda_{+}\int\rd^{4}\rho_{+}~~f(x,\zeta+\lambda_{+},\brzeta)\exp\left(-\frac{\partial~}{\partial\lambda_{+}^{\alpha}}\frac{\partial~}{\partial \brzeta_{\alpha}}\right)\left[e^{-\brrho_{+}\lambda_{+}}g(x,\zeta,\brzeta)\right]}\\
{}=\textstyle{\frac{1}{(2\pi)^{4}}}\dis{\int\rd^{4} \lambda_{+}\int\rd^{4}\rho_{+}~~f(x,\zeta+\lambda_{+},\brzeta)\exp\left(\frac{\lpartial~}{\partial\lambda_{+}^{\alpha}}\frac{\rpartial~}{\partial \brzeta_{\alpha}}\right)\left[e^{-\brrho_{+}\lambda_{+}}g(x,\zeta,\brzeta)\right]}\\
{}=\textstyle{\frac{1}{(2\pi)^{4}}}\dis{\int\rd^{4} \lambda_{+}\int\rd^{4}\rho_{+}~~f(x,\zeta+\lambda_{+},\brzeta)\exp\left(\frac{\lpartial~}{\partial\zeta^{\alpha}}\frac{\rpartial~}{\partial \brzeta_{\alpha}}\right)\left[e^{-\brrho_{+}\lambda_{+}}g(x,\zeta,\brzeta)\right]}\\
{}=\textstyle{\frac{1}{(2\pi)^{4}}}\dis{\int\rd^{4} \lambda_{+}\int\rd^{4}\rho_{+}~~e^{\bar{\lambda}_{+}\rho_{+}}f(x,\zeta+\lambda_{+},\brzeta)\exp\left(\frac{\lpartial~}{\partial\zeta^{\alpha}}\frac{\rpartial~}{\partial \brzeta_{\alpha}}\right)g(z,\zeta,\brzeta)}\\
{}=\dis{\int\rd^{4}\lambda_{+}~~\delta(\lambda_{+})\,f(x,\zeta+\lambda_{+},\brzeta)\exp\left(\frac{\lpartial~}{\partial\zeta^{\alpha}}\frac{\rpartial~}{\partial \brzeta_{\alpha}}\right)g(x,\zeta,\brzeta)}\\
{}=\dis{f(x,\zeta,\brzeta)\exp\left({\frac{\lpartial~\,}{\partial \zeta^{\alpha}}\frac{\rpartial~}{\partial \brzeta_{\alpha}}}\right)g(x,\zeta,\brzeta)}\,,
\ea
\label{Aequivalence}
\ee
where, to proceed     from the third  line to the fourth line,  integrations  by part have been performed.\\

\subsection{Associativity of the star product~\eqref{associativity}\label{Appendixassociativity}}

The star product satisfies the associativity~(\ref{associativity}) as
\be
\ba{l}
\left(f\star g\right)\star h\\
=\textstyle{\frac{1}{(2\pi)^{4}}}\dis{\int\rd^{4} \lambda_{+}\int\rd^{4}\rho_{+}~~e^{\bar{\lambda}_{+}\rho_{+}}\left[f(\zeta+\lambda_{+},\brzeta)\star g(\zeta+\lambda_{+},\brzeta)\right] h(\zeta,\brzeta+\bar{\rho}_{+})}\\
=\textstyle{\frac{1}{(2\pi)^{8}}}\dis{\int\rd^{4} \lambda_{+}\int\rd^{4}\rho_{+}\int\rd^{4} \lambda^{\prime}_{+}\int\rd^{4}\rho^{\prime}_{+}~
e^{\bar{\lambda}_{+}\rho_{+}+\bar{\lambda}^{\prime}_{+}\rho^{\prime}_{+}}f(\zeta+\lambda_{+}+\lambda^{\prime}_{+},\brzeta) g(\zeta+\lambda_{+},\brzeta+\brrho^{\prime}_{+}) 
h(\zeta,\brzeta+\bar{\rho}_{+})}\\
=\textstyle{\frac{1}{(2\pi)^{8}}}\dis{\int\rd^{4} \lambda_{+}\int\rd^{4}\rho^{\prime\prime}_{+}\int\rd^{4} \lambda^{\prime\prime}_{+}\int\rd^{4}\rho^{\prime}_{+}~
e^{\bar{\lambda}_{+}\rho^{\prime\prime}_{+}+\bar{\lambda}^{\prime\prime}_{+}\rho^{\prime}_{+}}
f(\zeta+\lambda^{\prime\prime}_{+},\brzeta) g(\zeta+\lambda_{+},\brzeta+\brrho^{\prime}_{+}) 
h(\zeta,\brzeta+\bar{\rho}^{\prime}_{+}+\bar{\rho}^{\prime\prime}_{+})}\\
=\textstyle{\frac{1}{(2\pi)^{4}}}\dis{\int\rd^{4} \lambda^{\prime\prime}_{+}\int\rd^{4}\rho^{\prime}_{+}~
e^{\bar{\lambda}^{\prime\prime}_{+}\rho^{\prime}_{+}}
f(\zeta+\lambda^{\prime\prime}_{+},\brzeta) \left[g(\zeta,\brzeta+\brrho^{\prime}_{+}) \star
h(\zeta,\brzeta+\bar{\rho}^{\prime}_{+})\right]}\\
=f\star\left( g\star h\right)\,,
\ea
\label{Aassociativity}
\ee
where we have made the  change of  variables for the Majorana spinors:  $\lambda_{+}^{\prime\prime}=\lambda_{+}+\lambda_{+}^{\prime}$ and $\rho_{+}^{\prime\prime}=\rho_{+}-\rho_{+}^{\prime}$.\\

\subsection{Isomorphism, \eqref{isomorphism} \label{Appendixisomorphism}}
Henceforth we prove the isomorphism~\eqref{isomorphism},\footnote{For the alternative Weyl normal ordered star product, see \textit{e.g.}~\cite{Seiberg:1999vs,Park:2003ku}.}
\be
\colon f(\hzeta,\hbrzeta)\colon\colon g(\hzeta,\hbrzeta)\colon\,=\,\hcO\!\left[f(\zeta,\brzeta)\star g(\zeta,\brzeta)\right]\,.
\ee
Any hatted  object is an operator. In particular, the bosonic spinorial coordinates, $\zeta^{\alpha}$ and  $\brzeta_{\beta}$, are mapped  to the  operators, 
\be
\ba{ll}
\hzeta^{\alpha}=\hcO\!\left[\zeta^{\alpha}\right]\,,\quad&\quad
\hbrzeta_{\alpha}=\hcO\!\left[\brzeta_{\alpha}\right]\,,
\ea
\ee
which satisfy the non-commutative algebra,  \textit{c.f.~}\eqref{zetacom},
\be 
\ba{lll}
{}\big[\hzeta^{\alpha}\,,\,\hbrzeta_{\beta}\big]=\hzeta^{\alpha}\hbrzeta_{\beta}-\hbrzeta_{\beta}\hzeta^{\alpha}=\delta^{\alpha}_{~\beta}\,,\quad&\quad
\hzeta^{\alpha}\hzeta^{\beta}=\hzeta^{\beta}\hzeta^{\alpha}\,,
\quad&\quad
\hbrzeta_{\alpha}\hbrzeta_{\beta}=\hbrzeta_{\beta}\hbrzeta_{\alpha}\,.
\ea
\label{zetacom2}
\ee
The Wick ordering, denoted by the colon, prescribes to  place all the unbarred  (annihilation) operators, $\hzeta^{\alpha}$,  to the right and    the barred  (creation)  operators, $\hbrzeta_{\beta}$, to the left. For example, 
\be
\ba{lll}
\colon\hbrzeta_{\beta}\hzeta^{\alpha}
\colon\,=\hbrzeta_{\beta}\hzeta^{\alpha}\,,\quad&\quad
\colon\hzeta^{\alpha}\hbrzeta_{\gamma}\hzeta^{\beta}
\colon\,=\hbrzeta_{\gamma}\hzeta^{\alpha}\hzeta^{\beta}\,,
\quad&\quad
\colon\hbrzeta_{\beta}\hzeta^{\alpha}\hbrzeta_{\delta}\hzeta^{\gamma}
\colon\,=\hbrzeta_{\beta}\hbrzeta_{\delta}\hzeta^{\alpha}\hzeta^{\gamma}\,.
\ea
\ee
For an arbitrary function of the internal commuting coordinates, $f(\zeta,\brzeta)$,  the corresponding  operator, $\hcO[f(\zeta,\brzeta)]$, is defined  subject to  the Wick   ordering  prescription, 
\be
\hcO\!\left[f(\zeta,\brzeta)\right]=\,\colon f(\hzeta,\hbrzeta) \colon\,.
\ee
It is straightforward to verify the following  preliminary relations to the isomorphism~(\ref{isomorphism}),
\be
\ba{ll}
\zeta^{\alpha}\star f(\zeta,\brzeta)=\zeta^{\alpha} f(\zeta,\brzeta)+\frac{\partial~}{\partial\brzeta_{\alpha}}f(x,\zeta,\brzeta)\,,\quad&\quad \brzeta_{\alpha}\star f(\zeta,\brzeta)=\brzeta_{\alpha} f(\zeta,\brzeta)\,,\\
f(\zeta,\brzeta)\star\brzeta_{\alpha}=\brzeta_{\alpha} f(\zeta,\brzeta)+\frac{\partial~}{\partial\zeta^{\alpha}}f(x,\zeta,\brzeta)\,,

\quad&\quad
f(\zeta,\brzeta)\star \zeta^{\alpha}=\zeta^{\alpha} f(\zeta,\brzeta)\,,
\ea
\ee
and
\be
\ba{l}
\hzeta^{\alpha}\hcO\!\left[f(\zeta,\brzeta)\right]=\hzeta^{\alpha}\colon f(\hzeta,\hbrzeta)\colon=
\colon\hzeta^{\alpha} f(\hzeta,\hbrzeta)\colon+\big[\,\hzeta^{\alpha}\,,\,\colon f(\hzeta,\hbrzeta)\colon\big]
=\hcO\!\left[\zeta^{\alpha}\star f(\zeta,\brzeta)\right]\,,\\
\hbrzeta_{\alpha}\hcO\!\left[f(\zeta,\brzeta)\right]=\hbrzeta_{\alpha}\colon f(\hzeta,\hbrzeta)\colon=\colon\hbrzeta_{\alpha}f(\hzeta,\hbrzeta)\colon=\hcO\!\left[\brzeta_{\alpha}f(\zeta,\brzeta)\right]=\hcO\!\left[\brzeta_{\alpha}\star f(\zeta,\brzeta)\right]\,.
\ea
\label{preliminary}
\ee
Now we assume that, the isomorphism~(\ref{isomorphism}) holds up to the $n\,$th order polynomials  of $\zeta$ and $\brzeta$, say $f_{n}(\zeta,\brzeta)$, and    an arbitrary function, $g(\zeta,\brzeta)$, \textit{i.e.}
\be
\colon f_{n}(\hzeta,\hbrzeta)\colon\colon g(\hzeta,\hbrzeta)\colon\,=\,\hcO\!\left[f_{n}(\zeta,\brzeta)\star g(\zeta,\brzeta)\right]\,.
\label{isomorphism2}
\ee
The preliminary results~(\ref{preliminary}) show that indeed (\ref{isomorphism2}) holds for $n=1$.  In order to establish an mathematical induction proof, we need to consider  $(n{+1})\,$th order polynomials,  or equivalently  both  $\zeta^{\alpha}f_{n}(\zeta,\brzeta)$ and   
 $\brzeta_{\alpha}f_{n}(\zeta,\brzeta)$. Utilizing  \eqref{preliminary}, \eqref{isomorphism2} and  the associativity of the  product, we get 
\be
\ba{ll}
\colon \hzeta^{\alpha}f_{n}(\hzeta,\hbrzeta)\colon\colon g(\hzeta,\hbrzeta)\colon\!\!\!\!\!&=\,
\colon f_{n}(\hzeta,\hbrzeta)\colon \hzeta^{\alpha}\colon g(\hzeta,\hbrzeta)\colon\\
{}&=\,
\colon f_{n}(\hzeta,\hbrzeta)\colon\hcO\!\left[ \zeta^{\alpha}\star g(\zeta,\brzeta)\right]\\
{}&=\hcO\!\left[f_{n}(\zeta,\brzeta)\star\left\{\zeta^{\alpha}\star g(\zeta,\brzeta)\right\}\right]\\
{}&=\hcO\!\left[\left\{f_{n}(\zeta,\brzeta)\star\zeta^{\alpha}\right\}\star g(\zeta,\brzeta)\right]\\
{}&=\hcO\!\left[\left\{\zeta^{\alpha}f_{n}(\zeta,\brzeta)\right\}\star g(\zeta,\brzeta)\right]\,,
\ea
\ee
and also 
\be
\ba{ll}
\colon \hbrzeta_{\alpha}f_{n}(\hzeta,\hbrzeta)\colon\colon g(\hzeta,\hbrzeta)\colon\!\!\!\!\!&=\,\hbrzeta_{\alpha}
\colon f_{n}(\hzeta,\hbrzeta)\colon \colon g(\hzeta,\hbrzeta)\colon\\
{}&=\,\hbrzeta_{\alpha}\,\hcO\!\left[f_{n}(\zeta,\brzeta)\star g(\zeta,\brzeta)\right]\\
{}&=\,\hcO\!\left[\brzeta_{\alpha}\star\left\{f_{n}(\zeta,\brzeta)\star g(\zeta,\brzeta)\right\}\right]\\
{}&=\,\hcO\!\left[\left\{\brzeta_{\alpha}\star f_{n}(\zeta,\brzeta)\right\}\star g(\zeta,\brzeta)\right]\\
{}&=\,\hcO\!\left[\left\{\brzeta_{\alpha} f_{n}(\zeta,\brzeta)\right\}\star g(\zeta,\brzeta)\right]\,.
\ea
\ee
These two results complete our mathematical induction proof.\\


\providecommand{\href}[2]{#2}\begingroup\raggedright

\end{document}